\documentclass[aps,physrev,reprint,superscriptaddress,longbibliography]{revtex4-2}
\usepackage{graphicx}
\usepackage{amsmath,amssymb}
\usepackage{quantikz}
\usepackage{hyperref}
\usepackage{xcolor}
\hypersetup{colorlinks=true,linkcolor=blue,urlcolor=blue,citecolor=blue}
\usepackage{orcidlink}
\usepackage{comment}

\begin{document}

\title{Problem-Specific Basis Quantum State Readout via Proper Orthogonal
Decomposition}
\author{Kota Ichiki\orcidlink{0009-0007-6015-377X}}
\thanks{Contact author: k.ichiki@scsk.jp}
\affiliation{Intellectual Property $\&$ Technology Strategy Division, SCSK Corporation, Koto, Tokyo 135-8110, Japan}

\author{Xinchi Huang\orcidlink{0000-0002-7547-074X}}
\affiliation{Quemix Inc., Taiyo Life Nihonbashi Building, 2-11-2, Nihonbashi Chuo-ku, Tokyo 103-0027, Japan}
\affiliation{Department of Physics, The University of Tokyo, Tokyo 113-0033, Japan}

\author{Gekko Budiutama\orcidlink{0000-0001-8904-9983}}
\affiliation{Quemix Inc., Taiyo Life Nihonbashi Building, 2-11-2, Nihonbashi Chuo-ku, Tokyo 103-0027, Japan}
\affiliation{Department of Physics, The University of Tokyo, Tokyo 113-0033, Japan}

\author{Masari Watanabe\orcidlink{0000-0003-3186-535X}}
\affiliation{Quemix Inc., Taiyo Life Nihonbashi Building, 2-11-2, Nihonbashi Chuo-ku, Tokyo 103-0027, Japan}
\affiliation{Department of Physics, The University of Tokyo, Tokyo 113-0033, Japan}

\author{Yoshifumi Kawada\orcidlink{0000-0002-1112-5330}}
\affiliation{Quemix Inc., Taiyo Life Nihonbashi Building, 2-11-2, Nihonbashi Chuo-ku, Tokyo 103-0027, Japan}
\affiliation{Department of Physics, The University of Tokyo, Tokyo 113-0033, Japan}

\author{Ryunosuke Terasawa\orcidlink{0009-0008-2504-8437}}
\affiliation{Quemix Inc., Taiyo Life Nihonbashi Building, 2-11-2, Nihonbashi Chuo-ku, Tokyo 103-0027, Japan}
\affiliation{Department of Physics, The University of Tokyo, Tokyo 113-0033, Japan}

\author{Hirofumi Nishi\orcidlink{0000-0001-5155-6605}}
\affiliation{Quemix Inc., Taiyo Life Nihonbashi Building, 2-11-2, Nihonbashi Chuo-ku, Tokyo 103-0027, Japan}
\affiliation{Department of Physics, The University of Tokyo, Tokyo 113-0033, Japan}

\author{Takayuki Suzuki\orcidlink{0000-0003-3400-976X}}
\affiliation{Intellectual Property $\&$ Technology Strategy Division, SCSK Corporation, Koto, Tokyo 135-8110, Japan}

\author{Ryutaro Nagai\orcidlink{0009-0000-4083-936X}}
\affiliation{Intellectual Property $\&$ Technology Strategy Division, SCSK Corporation, Koto, Tokyo 135-8110, Japan}

\author{Yu-ichiro Matsushita\orcidlink{0000-0002-9254-5918}}
\affiliation{Quemix Inc., Taiyo Life Nihonbashi Building, 2-11-2, Nihonbashi Chuo-ku, Tokyo 103-0027, Japan}
\affiliation{Department of Physics, The University of Tokyo, Tokyo 113-0033, Japan}
\affiliation{Quantum Materials and Applications Research Center, National Institutes for Quantum Science and Technology, Tokyo 152-8550, Japan}

\begin{abstract}
Quantum computing offers a promising approach to accelerating partial differential equation (PDE) solvers for large-scale, real-world problems. However, reconstructing the classical representation of a solution from a quantum state remains a significant computational bottleneck. We propose a problem-specific method, termed proper orthogonal decomposition-based readout (PODR), to improve this efficiency. This method comprises offline and online stages. In the offline stage, a set of basis functions capturing the dominant features of the target problem is constructed using classical computations. In the online stage, the quantum state is projected onto this reduced basis, and only a small number of coefficients is extracted to reconstruct the solution. PODR is particularly advantageous for simulations involving varying parameters—common in computational fluid dynamics (CFD)—because the POD basis functions are constructed only once in the offline stage and reused during the online stage. Applying PODR to benchmark CFD problems demonstrates a significant reduction in online computational cost compared with conventional readout methods.
\end{abstract}

\maketitle

\section{Introduction}

Quantum computing has the potential to accelerate the solution of partial differential equations (PDEs) arising in large-scale, real-world applications. 
A variety of quantum algorithms have been proposed for this purpose, including quantum linear systems algorithms (QLSAs) \cite{PhysRevLett.103.150502,ambainis:LIPIcs.STACS.2012.636,Berry:2014jeh,Berry:2017fxp,AndrewM:2015yvx,Childs:2019ddv,Lin2020optimalpolynomial,Childs2021highprecision,PhysRevA.110.012422,Costa:2021viq,PhysRevLett.110.250504,Bagherimehrab:2023mvu}, probabilistic imaginary-time evolution \cite{Huang:2024ixj}, and Schr\"{o}dingerization \cite{Huang:2024ixj,PhysRevLett.131.150603,PhysRevResearch.5.043220,Kieferova:2018kmn,Fang2023timemarchingbased,PhysRevA.108.032603,PhysRevLett.133.230602,PhysRevResearch.6.033246,PhysRevApplied.23.014063}. 
These approaches have attracted considerable attention because of their potential applicability to practical fields such as computer-aided engineering (CAE) and computational fluid dynamics (CFD) \cite{Kadowaki:2025mvzAI,Tucker:2025krr,Luckow2021}.

Although quantum algorithms for PDEs can provide favorable scaling with respect to the grid size, extracting useful classical information from the resulting quantum state remains a major bottleneck. 
Standard readout approaches include quantum state tomography \cite{Vogel:1989zz} and quantum amplitude estimation (QAE) \cite{Brassard:2000xvp,Manzano2023RealQAE,quantum6010001}. 
In this paper, we refer to the simplest readout strategy, in which all qubits are measured in the computational basis without using QAE, as real-space readout (RSR). 
Unless the solution state has sparse support \cite{Chen:2024eip}, the number of measurements required by RSR scales with the grid size $N$, which can severely limit its applicability to large-scale problems.

Basis-function-based readout has recently emerged as a promising strategy for reducing this dependence on $N$. 
The central idea is to estimate a small number of expansion coefficients from the quantum state and then reconstruct the solution classically from these coefficients. 
For example, a nonorthogonal basis of Lorentzian functions was used in Ref.~\cite{Nishi:2025ffn} for a specific class of functions, exploiting the efficient encoding of Lorentzian functions \cite{Kosugi:2024eus,Kosugi:2025lzw}. 
Orthogonal bases, including trigonometric functions and Chebyshev polynomials, have also been proposed for general real-valued functions \cite{article,Su:2025wtt,Williams:2023cuz,Fang:2025jux,Huang_2026,Huang:2025exw}. 
In particular, Fourier space readout (FSR) \cite{Huang_2026,Huang:2025exw} can identify dominant basis functions without requiring prior knowledge of the solution and can achieve scaling independent of $N$, although the number of retained Fourier basis functions generally depends on the error tolerance and the target solution.
Thus, basis-function-based methods can substantially reduce the readout cost associated with $N$. However, fixed bases are not necessarily well adapted to all solution manifolds. In FSR, for example, the Fourier basis may be inefficient for representing solutions with localized structures or discontinuities. Identifying basis functions that are well matched to the target problem therefore remains an important challenge in quantum state readout.

In this work, we address this challenge by introducing a proper-orthogonal-decomposition-based readout method, which we call PODR; see Fig.~\ref{fig:frog}. 
The method follows the offline-online decomposition paradigm widely used in reduced-order modeling in the CAE and CFD communities \cite{Rozza2008RB,doi:10.2514/1.J056060,RowleyDawson2017ModelReduction,Benner2015ASO,doi:10.1137/19M1257275,Ostrowski01012008,10.1108/HFF-03-2016-0083,cmes.2020.08164,DEHGHAN2017478,Rogers01072012,DUTTA2021110378}. 
In the offline stage, which is performed once and may scale with $N$, proper orthogonal decomposition (POD) \cite{BerkoozHolmesLumley1993POD} is used to construct problem-specific basis functions from representative solution snapshots. We refer to these basis functions as POD basis functions.
In the online stage, the solution state is read out by estimating only the coefficients associated with these POD basis functions, after which the solution is reconstructed on a classical computer. 
Because the POD basis functions are adapted to the solution manifold of the target problem, accurate reconstruction can often be achieved with only a small number of basis functions.

We also provide a systematic analysis of the classical and quantum computational resources required by PODR and compare them with those of existing readout methods; see Table~\ref{tab:rsr_fsr_pod_complexity}. 
Let $\varepsilon$ denote the error tolerance in the reconstructed solution.
Under the scaling assumptions specified in Sec.~\ref{subsec:gate}, the quantum complexity of the online stage is
$\mathcal{O}\!\left((1/\varepsilon)^2 \mathrm{polylog}(N,1/\varepsilon)\right)$.
This scaling indicates that, once POD basis functions have been constructed offline, the online readout cost depends only polylogarithmically on the grid size $N$, apart from the measurement precision required for coefficient estimation. 
We further demonstrate the performance of PODR by reconstructing solutions for both steady-state and unsteady CFD benchmark problems; see Figs.~\ref{fig:cavity} and \ref{fig:karmann}. 
The results show that a single offline stage can produce problem-specific basis functions that substantially improve readout accuracy in practical problems while requiring only a limited number of measurements.

The remainder of this paper is organized as follows. 
Section~\ref{shuhou} presents the PODR algorithm and analyzes its computational cost. 
Section~\ref{meth} describes the experimental settings used to validate the proposed method. 
Section~\ref{cavityeval} presents the validation results and compares PODR with FSR and RSR. 
Finally, Sec.~\ref{conc} summarizes our findings and discusses future directions.

\section{POD Readout}
\label{shuhou}

\begin{figure*}[t]
\centering
\includegraphics[width=0.95\linewidth]{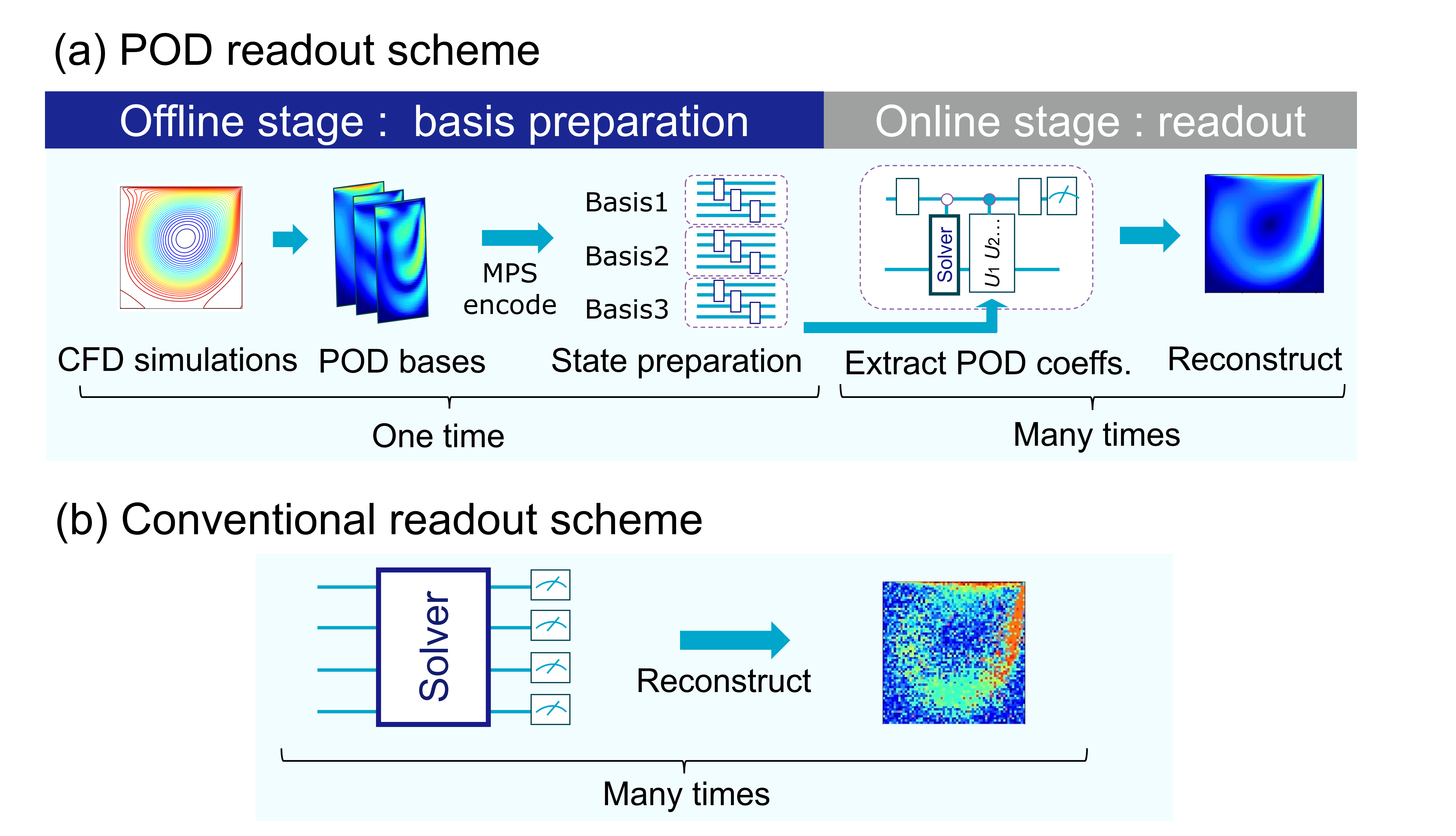}
\caption{\label{fig:frog}
(a) Workflow of the proposed proper orthogonal decomposition-based readout (PODR) method.
In the offline stage, POD basis functions are constructed from classical CAE/CFD simulation data and encoded into quantum circuits via matrix product states (MPS).
This stage is executed once and may scale with $N$.
In the online stage, the solution state is prepared by a quantum CAE/CFD solver, whose computational complexity is assumed to scale as $\mathcal{O}(\mathrm{polylog}(N))$~\cite{Huang:2024ixj,PhysRevLett.131.150603,PhysRevResearch.5.043220,Kieferova:2018kmn,Fang2023timemarchingbased,PhysRevA.108.032603,PhysRevLett.133.230602,PhysRevResearch.6.033246,PhysRevApplied.23.014063,Huang:2024ogm}.
The overlaps between the solution state and the approximated POD basis functions, $\langle \boldsymbol{x} \mid \tilde{\boldsymbol{u}}_i \rangle$, are then estimated, and the solution is reconstructed from the resulting coefficients.
Since the POD basis functions are prepared only once in the offline stage, they can be reused when the physical parameters are changed, enabling efficient repeated online readout.
(b) Schematic illustration of a conventional quantum readout method.
To estimate the solution at each grid point, the required number of shots scales with $N$, making accurate reconstruction difficult when the number of measurements is limited.}
\end{figure*}

Proper orthogonal decomposition-based readout (PODR) is a quantum state readout method that consists of two stages: an offline stage and an online stage (Fig.~\ref{fig:frog}(a)), in contrast to conventional readout schemes (Fig.~\ref{fig:frog}(b)). In the offline stage, POD basis functions are constructed from classical simulation data and encoded into quantum circuits. In the online stage, the coefficients associated with these basis functions are estimated via the Hadamard test, enabling reconstruction of the solution on a classical computer.

\subsection{Offline stage (basis preparation)}
\label{offline}
First, classical CFD simulations are performed while varying the external conditions of the target system or time steps, corresponding to the “CFD simulations” shown in Fig.~\ref{fig:frog}(a). 
In this study, we sample $M$ steady-state solutions obtained under different external conditions or $M$ time-step solutions of an unsteady flow under fixed external conditions (see Sec.~\ref{meth:nshot} for details of the experimental settings). Here, we assume that the number of samples $M$ is much smaller than $N$ $(M \ll N)$. 

Next, the sampled solutions are converted into one-dimensional vectors in row-major order and normalized. Each vector is defined as a snapshot $\boldsymbol{x}_i \in \mathbb{R}^N$, where $i$ indexes the corresponding parameter setting. Using these snapshots, a snapshot matrix is constructed as
\begin{equation}
\boldsymbol{S} =
\bigl[
\boldsymbol{x}_1\;\; \boldsymbol{x}_2\;\; \cdots\;\; \boldsymbol{x}_M
\bigr].
\end{equation}

Subsequently, to extract the dominant features shared by the $M$ sampled solutions, singular value decomposition (SVD) is applied to $\boldsymbol{S}$ as
\begin{eqnarray}
\label{tokuiti}
\boldsymbol{S} &=& \boldsymbol{U} \boldsymbol{\Sigma} \boldsymbol{V}^{\top}, \nonumber \\ 
\boldsymbol{U} &=& \bigl[\boldsymbol{u}_1\;\; \boldsymbol{u}_2\;\; \cdots\;\; \boldsymbol{u}_M\bigr], \nonumber \\
\boldsymbol{V} &=& \bigl[\boldsymbol{v}_1\;\; \boldsymbol{v}_2\;\; \cdots\;\; \boldsymbol{v}_M\bigr],
\end{eqnarray}
and the POD basis functions are obtained (corresponding to “POD basis functions” in Fig.~\ref{fig:frog}(a)). Here, $\boldsymbol{\Sigma} \in \mathbb{R}^{M \times M}$ is a diagonal matrix whose diagonal entries are the singular values. The vectors $\boldsymbol{u}_i \in \mathbb{R}^{N}$ and $\boldsymbol{v}_i \in \mathbb{R}^{M}$ denote the left and right singular vectors, respectively, ordered by decreasing singular values. Each vector $\boldsymbol{u}_i$ represents a characteristic spatial structure extracted from the $M$ sampled solutions and is referred to as a POD basis function. Since the basis functions are ordered by decreasing singular values, $\boldsymbol{u}_1$ corresponds to the most dominant spatial structure, and the contribution decreases as $i$ increases.

In the next step, to reduce the computational cost of the quantum readout, basis functions with small contributions are truncated, and the solution is approximated using only the leading $n_{\mathrm{b}}$ basis functions.
To select $n_{\mathrm{b}}$ for a prescribed error tolerance, the following estimated error is used:
\begin{equation}
\label{projest}
E^{\mathrm{est}}_{\mathrm{proj}} =
\sqrt{
\frac{1}{M}
\sum_{i=n_{\mathrm{b}}+1}^{M}
\sigma_i^2
}.
\end{equation}
In Eq.~(\ref{projest}), $\sigma_i$ denotes the $i$th singular value. A detailed derivation is provided in Appendix~\ref{estdousitu}. The value of $E^{\mathrm{est}}_{\mathrm{proj}}$ is evaluated for each $n_{\mathrm{b}}$, and the $n_{\mathrm{b}}$ that satisfies the prescribed error tolerance is selected. This determines the number of basis functions used for the readout.

Next, to efficiently encode the $n_{\mathrm{b}}$ POD basis functions into a quantum circuit, they are decomposed into matrix product states (MPS). According to previous studies~\cite{doi:10.1137/090752286,OSELEDETS201070}, once the maximum bond dimensions $\{ \chi_i\}_{i=1}^{n_{\mathrm{b}}}$ are specified, the normalized POD basis functions approximated by MPS, $\{\tilde{\boldsymbol{u}}_i\}_{i=1}^{n_{\mathrm{b}}}$, can be obtained.
Increasing the maximum bond dimensions improves the approximation accuracy; however, it also increases the implementation cost of the quantum circuit, as discussed later. Therefore, to determine $\{ \chi_i\}_{i=1}^{n_{\mathrm{b}}}$ that achieves the desired error tolerance, we evaluate the following estimated encoding error for each choice of bond dimensions, using the corresponding $\{\tilde{\boldsymbol{u}}_i\}_{i=1}^{n_{\mathrm{b}}}$:
\begin{equation}
\label{encest}
E^{\mathrm{est}}_{\mathrm{enc}}
=\sqrt{\sum_{i=1}^{n_{\mathrm{b}}} \left|
 \sqrt{\frac{\sigma_i^{2}}{M}}- \sum_{j=1}^{n_{\mathrm{b}}} \sqrt{\frac{\sigma_j^{2}}{M}}\, \tilde{\boldsymbol{u}}_i \cdot \boldsymbol{u}_j \right|^2} .
\end{equation}
A detailed derivation is provided in Appendix~\ref{estdousitu}. The set $\{ \chi_i\}_{i=1}^{n_{\mathrm{b}}}$ is varied within the range specified in Appendix~\ref{app:mps}, and the set of maximum bond dimensions is selected such that the estimated error $E^{\mathrm{est}}_{\mathrm{enc}}$ satisfies the prescribed error tolerance. The resulting $\{ \chi_i\}_{i=1}^{n_{\mathrm{b}}}$ and $\{\tilde{\boldsymbol{u}}_i\}_{i=1}^{n_{\mathrm{b}}}$ are then used in the final encoding step.

The approximated POD basis functions $\tilde{\boldsymbol{u}}_i$ obtained in this manner are then encoded into quantum circuits, corresponding to ``State preparation'' in Fig.~\ref{fig:frog}(a). Using the MPS-based state-preparation method~\cite{PRXQuantum.2.010342,SciPostPhys.14.5.117}, each approximated POD basis function can be represented by a unitary operator $U_i$ such that
$U_i \ket{0}^{\otimes n} = \ket{\tilde{\boldsymbol{u}}_i}$.
Accordingly, the selected POD basis functions are implemented as a set of state-preparation unitaries $\{U_i\}_{i=1}^{n_\mathrm{b}}$. The circuit shown in ``State preparation'' in Fig.~\ref{fig:frog}(a) illustrates an example of this encoding procedure for a grid size of 16 and a maximum bond dimension of 2.

\subsection{Online stage (readout)}
\label{sub:online}
Suppose that the solution has already been prepared as a quantum state $\ket{\boldsymbol{x}}$ by a quantum CAE/CFD solver. 
The main objective of PODR is to read out the solution by expressing $\ket{\boldsymbol{x}}$ as a linear combination of precomputed POD basis functions and estimating the corresponding coefficients:
\begin{eqnarray}
\ket{\boldsymbol{x}} \approx \sum_{i=1}^{n_\mathrm{b}} c_i \, \ket{\boldsymbol{u}_i},
\end{eqnarray}
where the expansion coefficient is given by $c_i = \langle \boldsymbol{u}_i \mid \boldsymbol{x} \rangle$. 

In practice, the corresponding coefficients are estimated approximately by measuring the overlap between the solution state $\ket{\boldsymbol{x}}$ and the approximated POD basis function $\ket{\tilde{\boldsymbol{u}}_i}$ prepared using the gate $U_i$.

In this study, the overlap is estimated using the Hadamard test, corresponding to “Extract POD coeffs.” in Fig.~\ref{fig:frog}(a). 
For a given $\ket{\tilde{\boldsymbol{u}}_i}$, let $p_{0,i}$ and $p_{1,i}$ denote the probabilities that the ancilla qubit is measured in the computational basis states $\ket{0}$ and $\ket{1}$, respectively. Then,
\begin{equation}
\label{6}
p_{0,i}
=
\frac{1 + \mathrm{Re}(\langle \boldsymbol{x} \mid \tilde{\boldsymbol{u}}_i\rangle )}{2}, \qquad
p_{1,i}
=
\frac{1 - \mathrm{Re}(\langle\boldsymbol{x} \mid \tilde{\boldsymbol{u}}_i \rangle ) } {2}.
\end{equation}
$\mathrm{Re}(\langle \boldsymbol{x} \mid \tilde{\boldsymbol{u}}_i\rangle ) =\langle \boldsymbol{x} \mid \tilde{\boldsymbol{u}}_i \rangle$ because $\boldsymbol{x}$ and $\tilde{\boldsymbol{u}}_i$ are real vectors, and $\mathrm{Re}(z)$ denotes the real part of $z$.
Let $Z_{0,i}$ and $Z_{1,i}$ denote the numbers of measurement outcomes corresponding to $\ket{0}$ and $\ket{1}$, respectively. Assuming that the total number of shots is distributed equally among the POD basis functions, the number of shots allocated to each basis function is given by
$N_{\mathrm{shot}}^{\mathrm{basis}} = Z_{0,i}+Z_{1,i}$.
The corresponding probabilities are then estimated as
\begin{equation}
\label{coeffeq}
\tilde{p}_{0,i}
=\frac{Z_{0,i}}{N_{\mathrm{shot}}^{\mathrm{basis}}}, \qquad
\tilde{p}_{1,i}
=\frac{Z_{1,i}}{N_{\mathrm{shot}}^{\mathrm{basis}}}.
\end{equation}
Accordingly, using $\langle\boldsymbol{x} \mid \tilde{\boldsymbol{u}}_i \rangle = 2p_{0,i}-1$ from Eq.~(\ref{6}), the coefficients associated with $\tilde{\boldsymbol{u}}_i$ are estimated as $\tilde{c}_i = 2 \tilde{p}_{0,i} - 1$, and the reconstructed solution is given by
\begin{equation}
\label{reconstructeq}
\tilde{\boldsymbol{x}}
=
\sum_{i=1}^{n_{\mathrm{b}}} \tilde{c}_i \boldsymbol{u}_i.
\end{equation}
This corresponds to “Reconstruct” in Fig.~\ref{fig:frog}(a). Note that the reconstruction is performed using the exact POD basis functions, rather than the approximated ones. If the region of interest is restricted to a subset of $J \le N$ grid points, the reconstruction can be carried out using only the corresponding components of $\boldsymbol{u}_i$.

\begin{table*}[t]
\centering
\caption{Classical and quantum resources for RSR, FSR, and PODR methods.}
\label{tab:rsr_fsr_pod_complexity}
\begin{tabular}{lcccc}
\hline
 & \multicolumn{1}{c}{\textbf{Offline}} & \ \ & \multicolumn{2}{c}{\textbf{Online}} \\
\cline{2-2}\cline{4-5}
Method
& Classical Complexity
& 
& Quantum Complexity
& Classical Complexity \\
\hline
RSR
& 0
&
& $\mathcal{O}\!\left( (1/\varepsilon)^2 N \mathrm{polylog}(N) \right)$
& $\mathcal{O}\!\left( N (1/\varepsilon)^2 \right)$ \\
FSR
& 0
&
& $\mathcal{O}\!\left((1/\varepsilon)^{2+s} \mathrm{polylog}(N) \right)$
& $\mathcal{O}\!\left( (1/\varepsilon)^{2+s} \right)
 + \mathcal{O}\!\left( J(1/\varepsilon)^s \right)$ \\
PODR
& $\mathcal{O}(N)+\mathcal{O}(Nn_\mathrm{b}\chi^2\mathrm{polylog}(\chi))$
&
& $\mathcal{O}\!\left(n_{\mathrm{b}}^2\chi^2(1/\varepsilon)^2\mathrm{polylog}(N)\right)$
& $\mathcal{O}\!\left( n_{\mathrm{b}}^2(1/\varepsilon)^2 \right) + \mathcal{O}\!\left( Jn_{\mathrm{b}} \right)$ \\
\hline
\end{tabular}
\end{table*}

\subsection{Readout error}
\label{readouterror}

In this analysis, we assume an ideal quantum computer and do not include errors arising from hardware noise, gate imperfections, or decoherence. 
The readout error of PODR is defined as
$\varepsilon = \|\boldsymbol{x} - \tilde{\boldsymbol{x}}\|_{2}$,
where $\boldsymbol{x}$ is the target solution and $\tilde{\boldsymbol{x}}$ is the reconstructed solution. 
This error can be bounded by three contributions:
\begin{align}
\label{leqq}
\varepsilon
&\le E_{\mathrm{proj}} + E_{\mathrm{enc}} + E_{\mathrm{sam}}, \\
E_{\mathrm{proj}} &= \| \boldsymbol{x} - \boldsymbol{U}_{n_{\mathrm{b}}} \boldsymbol{U}_{n_{\mathrm{b}}}^{\top} \boldsymbol{x} \|_{2}, \nonumber \\
E_{\mathrm{enc}} &= \sqrt{\sum_{i=1}^{n_{\mathrm{b}}} \left\lvert \langle \boldsymbol{x} \mid \boldsymbol{\delta}_i \rangle \right\rvert^2}, \nonumber \\
E_{\mathrm{sam}} &= \beta \sqrt{\frac{n_{\mathrm{b}}}{N_{\mathrm{shot}}^{\mathrm{basis}}}}. \nonumber
\end{align}
A detailed derivation is provided in Appendix~\ref{app:simeq}.

The first term, $E_{\mathrm{proj}}$, is defined as the error caused by reconstructing the solution using only the selected $n_{\mathrm{b}}$ POD basis functions. Here,
$\boldsymbol{U}_{n_{\mathrm{b}}}
= \bigl[\boldsymbol{u}_1\;\; \boldsymbol{u}_2\;\; \cdots\;\; \boldsymbol{u}_{n_{\mathrm{b}}}\bigr]$
denotes the matrix whose columns are the $n_{\mathrm{b}}$ POD basis functions used for readout.
Thus, $E_{\mathrm{proj}}$ measures the component of the solution $\boldsymbol{x}$ that cannot be represented in the subspace spanned by these POD basis functions.

The second term, $E_{\mathrm{enc}}$, is defined as the error caused by approximating each POD basis function by an MPS. For the $i$th POD basis function, let $\ket{\tilde{\boldsymbol{u}}_i}$ be its normalized MPS approximation, and define
$\ket{\boldsymbol{\delta}_i}
= \ket{\boldsymbol{u}_i} - \ket{\tilde{\boldsymbol{u}}_i}$.
In Eq.~(\ref{leqq}) $E_{\mathrm{enc}}$ evaluates how this MPS approximation error affects the readout result for the solution $\boldsymbol{x}$.

The third term, $E_{\mathrm{sam}}$, is the sampling error arising from the finite number of measurements used to estimate the POD coefficients via the Hadamard test. Here, $N_{\mathrm{shot}}^{\mathrm{basis}}$ denotes the number of shots allocated to each POD basis function, and $\beta$ is a sufficiently large constant; in this work, we take $\beta \ge 2$.

We note that the readout error quantities $E_{\mathrm{proj}}$ and $E_{\mathrm{enc}}$ introduced above should be interpreted with care. The definitions of $E_{\mathrm{proj}}$ and $E_{\mathrm{enc}}$ involve the true solution $\boldsymbol{x}$; therefore, these quantities cannot be evaluated exactly in practical settings unless $\boldsymbol{x}$ is available. In the following discussion, we assume that the true solution $\boldsymbol{x}$ is available in order to evaluate the theoretical readout error of PODR. In practical applications of PODR, however, the true solution $\boldsymbol{x}$ is unknown. In such cases, it is more realistic to use the estimated errors $E_{\mathrm{proj}}^{\mathrm{est}}$ and $E_{\mathrm{enc}}^{\mathrm{est}}$ introduced in Sec.~\ref{offline}.

\subsection{Computational complexity}
\label{meth:complexity}

We analyze the computational complexity of PODR with respect to $N$, $\varepsilon$, $J$, $n_{\mathrm{b}}$, and $\chi$, where $\chi \equiv \max_{1\le i \le n_{\mathrm{b}}}\chi_i$ and $\chi_i$ denotes the maximum bond dimension used to approximate the $i$th POD basis function. We divide the analysis into three parts: the classical complexity of the offline stage, the quantum complexity of the online stage, and the classical complexity of the online stage. We then compare the computational cost of PODR with those of RSR and FSR~\cite{Huang_2026,Huang:2025exw}, as summarized in Table~\ref{tab:rsr_fsr_pod_complexity}. In the complexity analysis of all methods, we assume that the quantum CAE/CFD solver preparing the solution state can be implemented with gate complexity $\mathcal{O}(\mathrm{polylog}(N))$~\cite{Huang:2024ixj,PhysRevLett.131.150603, PhysRevResearch.5.043220,Kieferova:2018kmn,Fang2023timemarchingbased, PhysRevA.108.032603,PhysRevLett.133.230602,PhysRevResearch.6.033246, PhysRevApplied.23.014063,Huang:2024ogm}.

For the offline stage of PODR, the classical complexity is dominated by
(i) computing the SVD of $\boldsymbol{S}$, which requires $\mathcal{O}(N)$ operations when $M \ll N$ is treated as fixed, and
(ii) constructing the MPS approximations and determining the bond dimensions
$\{\chi_i\}_{i=1}^{n_{\mathrm{b}}}$, whose computational cost is
$\mathcal{O}(Nn_\mathrm{b}\chi^2\log_2\chi)$.
Further details of the estimate in (ii) are provided in Appendix~\ref{app:mps}.
Therefore, the overall offline classical complexity is
$\mathcal{O}(N)+\mathcal{O}(Nn_\mathrm{b}\chi^2\mathrm{polylog}(\chi))$.

The quantum complexity of the online stage is governed by the extraction of the POD coefficients. We denote by $C_U$ an upper bound on the gate complexity of one circuit execution for estimating a POD coefficient.
This includes the preparation of both the solution state by the quantum CAE/CFD solver and the corresponding POD basis function.
With $N_{\mathrm{shot}}$ denoting the total number of shots, the online quantum complexity is $\mathcal{O}\!\left(C_U N_{\mathrm{shot}}\right)$.
We first estimate $C_U$. For each selected POD basis function, the corresponding MPS with maximum bond dimension $\chi$ is implemented with gate complexity $\mathcal{O}(\chi^2\log_2 N)$. The detailed derivation of this gate complexity is provided in Ref.~\cite{SciPostPhys.14.5.117}. Combining this with the assumed oracle cost of the quantum CAE/CFD solver $\mathcal{O}(\mathrm{polylog}(N))$, $C_U$ is bounded as
\begin{eqnarray}
\label{podgatecomp}
C_U = \mathcal{O}(\chi^2 \mathrm{polylog}(N)).
\end{eqnarray}

We next evaluate $N_{\mathrm{shot}}$. This quantity can be expressed as the product of $n_{\mathrm{b}}$ and the number of shots per basis function, $N_{\mathrm{shot}}^{\mathrm{basis}}$. From the sampling error bound in Eq.~(\ref{leqq}), the condition $E_{\mathrm{sam}} = \mathcal{O}(\varepsilon)$ gives
$N_{\mathrm{shot}}^{\mathrm{basis}} = \mathcal{O}(n_{\mathrm{b}}(1/\varepsilon)^2)$.
Since $N_{\mathrm{shot}} = n_{\mathrm{b}}N_{\mathrm{shot}}^{\mathrm{basis}}$, we obtain
\begin{eqnarray}
\label{shotscale}
N_{\mathrm{shot}} = \mathcal{O}\!\left(n_{\mathrm{b}}^2(1/\varepsilon)^2\right).
\end{eqnarray}
Accordingly, the quantum complexity of PODR is $\mathcal{O}\!\left(n_{\mathrm{b}}^2\chi^2(1/\varepsilon)^2\mathrm{polylog}(N)\right)$.

The classical complexity of the online stage is determined by the post-processing cost required to reconstruct the solution from the measurement outcomes. This cost consists of two parts: (i) estimating the POD coefficients from the measurement outcomes, which costs $\mathcal{O}(N_{\mathrm{shot}})= \mathcal{O}\!\left(n_{\mathrm{b}}^2(1/\varepsilon)^2\right)$, and (ii) evaluating the reconstructed solution at the $J$ target grid points by taking the corresponding linear combination of the POD basis function values, which costs $\mathcal{O}(J\,n_{\mathrm{b}})$. The resulting classical complexity of the online stage is $\mathcal{O}\!\left(n_{\mathrm{b}}^2(1/\varepsilon)^2\right)+ \mathcal{O}\!\left(Jn_{\mathrm{b}}\right)$.

We briefly summarize the computational cost of RSR~\cite{Huang_2026,Huang:2025exw}. RSR requires no $N$-dependent classical preprocessing, and thus incurs zero offline cost. In the online stage, the quantum complexity is determined solely by the number of shots $N_{\mathrm{shot}}$, which scales as $\mathcal{O}(N(1/\varepsilon)^2)$. Since each shot requires preparing the solution state, whose gate complexity is assumed to scale as
$\mathcal{O}(\mathrm{polylog}(N))$, the online quantum complexity of RSR is $\mathcal{O}\!\left(N(1/\varepsilon)^2\mathrm{polylog}(N)\right)$.
The classical complexity in the online stage corresponds to constructing a histogram from the measurement outcomes and is therefore of the same order as $N_{\mathrm{shot}}$.

We next summarize the computational cost of FSR~\cite{Huang_2026,Huang:2025exw}. As in RSR, no $N$-dependent classical preprocessing is required, and the offline cost is zero. In the online stage, the quantum complexity is given by the product of the gate complexity and the number of shots. Previous work has shown that this scales as $\mathcal{O}\!\left((1/\varepsilon)^{2+s}\mathrm{polylog}(N)\right)$, where $0 < s \leq 2$ characterizes the smoothness of the solution state. The classical complexity in the online stage consists of the cost of constructing a histogram of Fourier coefficients and evaluating the values at the target points $J$, yielding $\mathcal{O}\!\left((1/\varepsilon)^{2+s}\right) + \mathcal{O}\!\left(J(1/\varepsilon)^s\right)$.

Table~\ref{tab:rsr_fsr_pod_complexity} highlights the trade-off of PODR: it introduces an $N$-dependent offline cost, while its online quantum cost depends only polylogarithmically on $N$, apart from the dependence on the number of selected POD basis functions, their bond dimensions, and the error tolerance.
This trade-off is particularly relevant in practical CAE/CFD simulations, where multiple computations are performed for varying physical parameters. Since the offline stage is executed only once, the precomputed POD basis functions can be reused for repeated readout in the online stage.

\section{Experimental Setup}
\label{meth}

\subsection{Benchmark problems}
\label{meth:nshot}
We apply PODR to two benchmark problems: a steady two-dimensional (2D) lid-driven cavity flow and a 2D K\'arm\'an vortex street. The detailed configurations are described below.

\subsubsection{Steady 2D lid-driven cavity flow}
\label{setcav}
We consider the velocity field $(u_x, u_y)$ of a steady incompressible 2D lid-driven cavity flow~\cite{GHIA1982387} on a uniform grid of size $N = 256^2$. The velocity is set to zero on the side and bottom walls, while the top lid moves with a prescribed constant horizontal velocity. The snapshot data are computed using OpenFOAM~\cite{JASAK200989}. The Reynolds number $Re$ is used as the varying physical parameter to generate the snapshot ensemble.
The snapshot matrix $\boldsymbol{S}$ is constructed from samples with $Re = 100k$, where $k = 1,\ldots,10$ ($M = 10$). As a representative target state for evaluating the readout, we select $Re = 950$, which lies between the sampled values $Re = 900$ and $1000$. The rationale for choosing $Re=950$ is discussed in Appendix~\ref{app:snapshot_effect}.

\subsubsection{2D K\'arm\'an vortex street}
\label{setkar}
We study the time-dependent velocity field $(u_x, u_y)$ of an incompressible 2D K\'arm\'an vortex street generated by flow past a circular cylinder at $Re = 100$~\cite{1734798}, using a uniform grid of size $N = 256 \times 128$. The time-step size is $\Delta t = 0.05$ in the nondimensional time units of the original dataset. As boundary conditions, a constant inflow velocity is imposed at the left boundary, while a natural outflow condition is applied at the right boundary. On the upper and lower boundaries, the velocity is fixed to zero. The Navier-Stokes equations are solved using PhiFlow~\cite{2020arXiv200107457H}. This flow exhibits periodic behavior with a period of approximately 50 time steps, corresponding to a time interval of $50\Delta t = 2.5$. The snapshot matrix $\boldsymbol{S}$ is constructed from the approximately periodic portion of the time-series dataset, using solutions at time steps $t = 600, 601, \ldots, 1200$ ($M = 601$). The solution at $t = 1273$ is selected as the target state $\ket{\boldsymbol{x}}$ for evaluating the readout. The rationale for choosing $t = 1273$ is discussed in Appendix~\ref{app:snapshot_effect}.

\begin{table}[t]
\centering
\caption{Selected PODR parameters, $n_\mathrm{b}$ and $\chi \equiv \max_{1\le i \le n_{\mathrm{b}}}\chi_i$ for the steady 2D lid-driven cavity flow.}
\label{tab:nshot_setting}
\begin{tabular}{ccccc}
\hline
Case & $N$ & Target & $n_{\mathrm{b}}$ & $\chi$ \\
\hline
Case~1 & $256^2$ & $u_x$ & 5 & 16 \\
       &         & $u_y$ & 5 & 8  \\
\hline
Case~2 & $32^2$  & $u_x$ & 8 & 16 \\
       &         & $u_y$ & 8 & 16 \\
       & $64^2$  & $u_x$ & 7 & 16 \\
       &         & $u_y$ & 7 & 16 \\
       & $128^2$ & $u_x$ & 8 & 16 \\
       &         & $u_y$ & 7 & 16 \\
       & $256^2$ & $u_x$ & 7 & 16 \\
       &         & $u_y$ & 7 & 16 \\
\hline
\end{tabular}
\end{table}

\begin{table}[t]
\centering
\caption{Selected PODR parameters, $n_\mathrm{b}$ and $\chi \equiv \max_{1\le i \le n_{\mathrm{b}}}\chi_i$ for the 2D K\'arm\'an vortex street.}
\label{tab:nshot_setting_kar}
\begin{tabular}{ccccc}
\hline
Case & $N$ & Target & $n_{\mathrm{b}}$ & $\chi$ \\
\hline
Case~1 & $256 \times 128$ & $u_x$ & 7 & 16 \\
       &         & $u_y$ & 11 & 16  \\
\hline
Case~2 & $256 \times 128$ & $u_x$ & 12 & 16 \\
       &         & $u_y$ & 20 & 16 \\
\hline
\end{tabular}
\end{table}

\subsection{Determination of the number of POD basis functions and bond dimension}
\label{meth:readoutset}

We consider two readout settings, labeled Case~1 and Case~2, to select the number of POD basis functions $n_{\mathrm{b}}$ and the maximum bond dimensions $\{\chi_i\}_{i=1}^{n_\mathrm{b}}$; these cases correspond to prescribed error levels of $\varepsilon = 10^{-2}$ and $\varepsilon = 2\times 10^{-3}$, respectively. Following the procedure described in Sec.~\ref{offline}, $n_{\mathrm{b}}$ and $\{\chi_i\}_{i=1}^{n_{\mathrm{b}}}$ are selected using the estimated projection error $E_{\mathrm{proj}}^{\mathrm{est}}$ and
the estimated encoding error $E_{\mathrm{enc}}^{\mathrm{est}}$. In this procedure, the estimated errors satisfy the corresponding prescribed error levels, with $E_{\mathrm{proj}}^{\mathrm{est}} + E_{\mathrm{enc}}^{\mathrm{est}} \le 10^{-2}$
for Case~1 and $E_{\mathrm{proj}}^{\mathrm{est}} + E_{\mathrm{enc}}^{\mathrm{est}} \le 2\times10^{-3}$ for Case~2. The resulting readout settings for the steady 2D lid-driven cavity flow and the 2D K\'arm\'an vortex street are summarized in Tables~\ref{tab:nshot_setting} and \ref{tab:nshot_setting_kar}, respectively.

\subsection{Stream function}
\label{meth:vis}
We visualize the readout results obtained using RSR, FSR~\cite{Huang_2026,Huang:2025exw}, and PODR for the problem settings introduced in Sec.~\ref{meth:nshot}. In addition to the velocity field $\boldsymbol{u} = (u_x, u_y)$, we visualize the stream function $\psi$ derived from $\boldsymbol{u}$. The problems considered in Sec.~\ref{meth:nshot} involve incompressible flows, for which a stream function can be defined. The stream function $\psi(x,y)$ satisfies
\begin{equation}
u_x = \frac{\partial \psi}{\partial y}, \qquad
u_y = -\frac{\partial \psi}{\partial x},
\end{equation}
and can be expressed as
\begin{equation}
\label{stream}
\psi(x,y)
=
\int^{y}_{y_{\mathrm{min}}} u_x(x,\eta)\,\mathrm{d}\eta + C(x),
\end{equation}
where $y_{\min}$ denotes the lower boundary of the domain and $C(x)$ is an integration constant. In both benchmark problems, the boundary conditions in Sec.~\ref{meth:nshot} give $u_y(x,y_{\min})=0$ at the lower boundary. Therefore, as shown in Appendix~\ref{streamf}, $C(x)$ can be regarded as an $x$-independent constant $C_0$, which only sets the reference level of $\psi$; it is set to zero in the visualizations.

\begin{figure*}[t]
\includegraphics[width=0.7\linewidth]{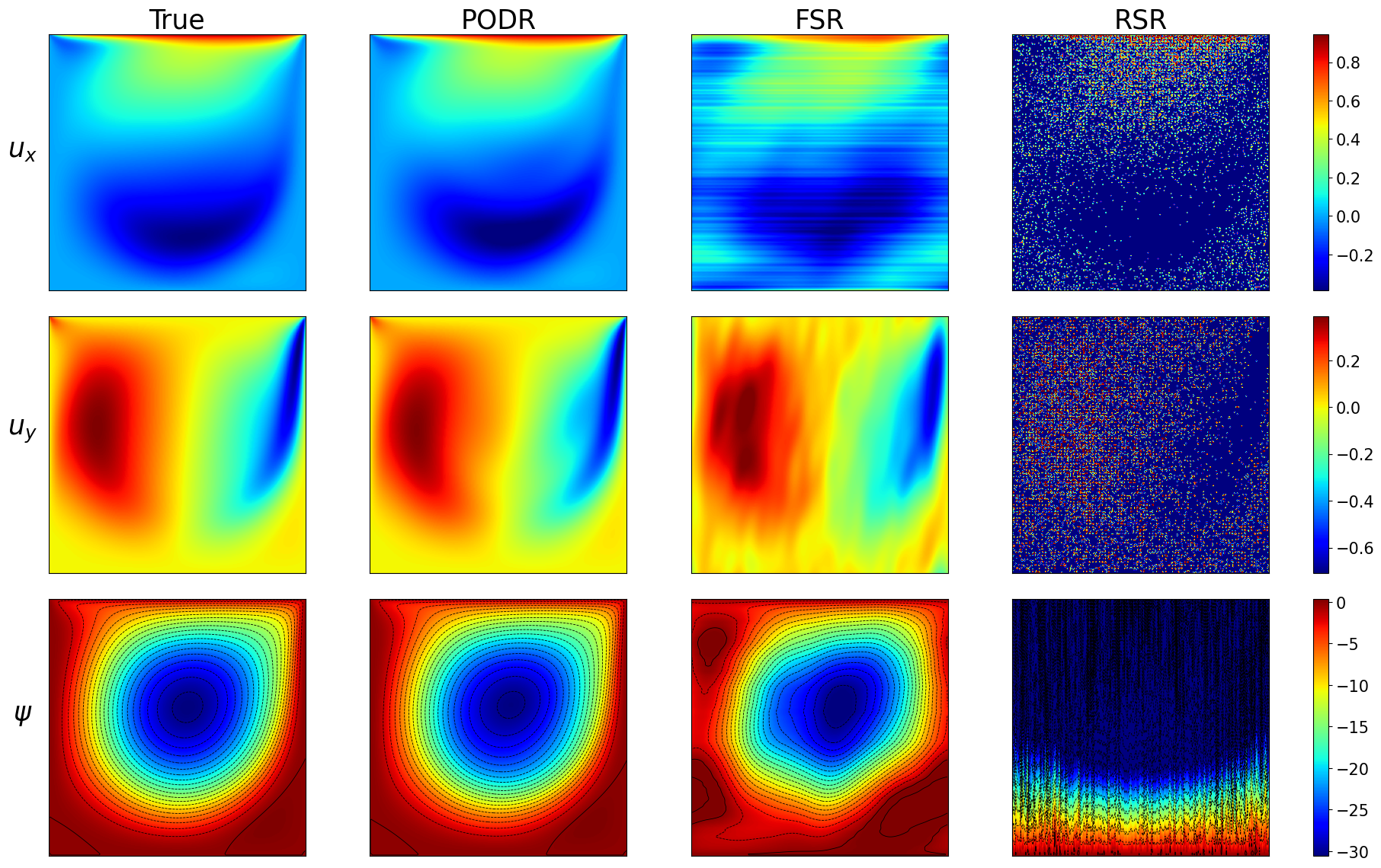}
\caption{\label{fig:cavity}Visualization of the reconstructed solution for the steady 2D lid-driven cavity flow at $N_{\mathrm{shot}} = 10^4$, comparing the readout methods. The parameter settings correspond to Case~1 for the $256^2$ grid listed in Table~\ref{tab:nshot_setting}.}

\includegraphics[width=0.95\linewidth]{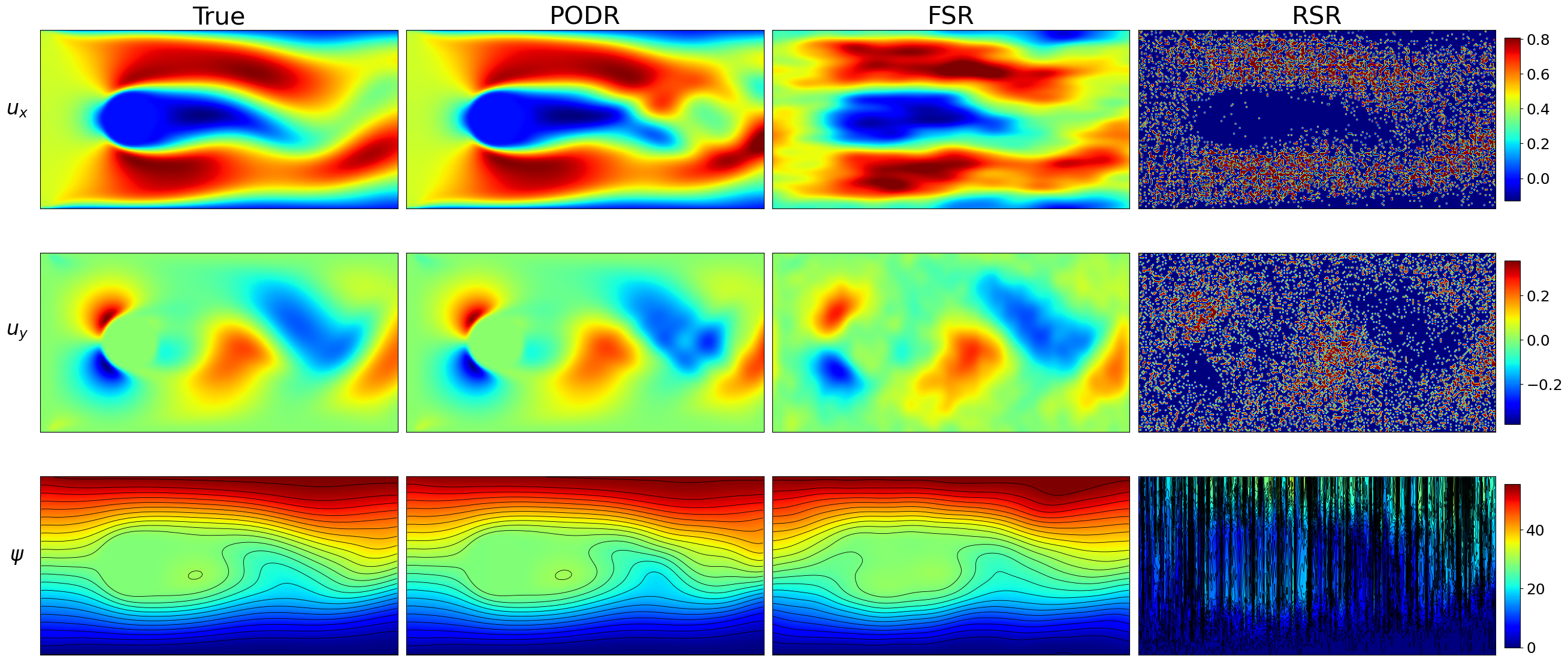}
\caption{\label{fig:karmann}Visualization of the reconstructed solution for the 2D K\'arm\'an vortex street at $N_{\mathrm{shot}} = 9933$, comparing the readout methods. The parameter settings correspond to Case~1 for the $256 \times 128$ grid listed in Table~\ref{tab:nshot_setting_kar}. The value \(N_{\mathrm{shot}}=9933\) results from allocating an equal integer number of shots to each selected POD basis function.}
\end{figure*}

\section{Results}
\label{cavityeval}

\subsection{Visualization of flow fields}
\label{visres}
Figs.~\ref{fig:cavity} and \ref{fig:karmann} compare the velocity fields reconstructed by the readout methods for the lid-driven cavity flow and the K\'arm\'an vortex street, respectively. At the selected values of $N_{\mathrm{shot}}$, RSR and FSR show visible reconstruction artifacts. RSR exhibits pointwise sampling noise in the velocity field, which is reflected in the stream function. FSR yields a smoother field by sampling Fourier coefficients, but discrepancies are still observed throughout the flow field. 

In contrast, PODR more closely reproduces the reference velocity field and stream-function contours under the same shot budget. For the lid-driven cavity flow, PODR reproduces the primary vortex structure and the corresponding stream-function contours. For the K\'arm\'an vortex street, PODR captures the alternating vortex pattern behind the cylinder and preserves the main wake structure observed in the reference solution. These results illustrate the effectiveness of using POD basis functions for visualization in the considered benchmarks.

\begin{figure}[t]

\raggedright
(a)\\
\includegraphics[width=0.95\linewidth]{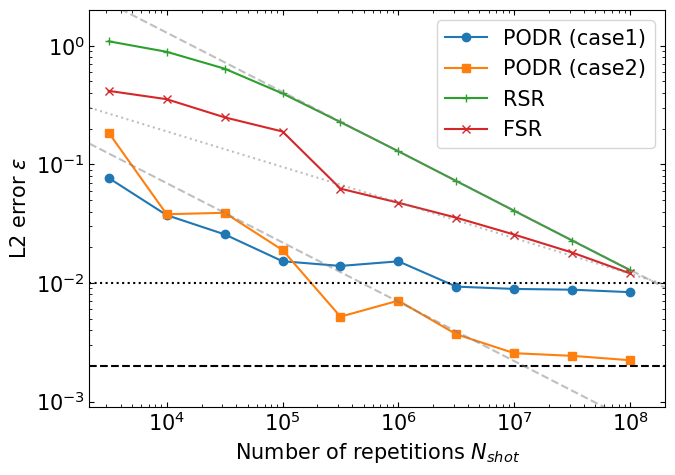}

\vspace{0.5em}

(b)\\
\includegraphics[width=0.95\linewidth]{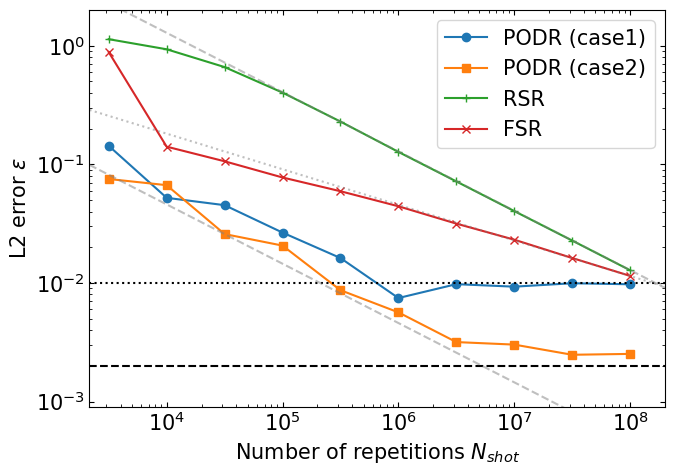}

\caption{\label{fig:2}
(a) Readout error as a function of the number of shots for the velocity field $u_x$ of the steady 2D lid-driven cavity flow.
The gray dashed lines indicate $\mathcal{O}(N_{\mathrm{shot}}^{-1/2})$ scaling for RSR and PODR, and the gray dotted line indicates $\mathcal{O}(N_{\mathrm{shot}}^{-3/10})$ scaling for FSR. The parameter settings used here are given by the Case~1 and Case~2 entries for the $256^2$ grid in Table~\ref{tab:nshot_setting}. The corresponding prescribed errors are indicated by a black dotted horizontal line for Case~1 and a black dashed horizontal line for Case~2. (b) Readout error $\varepsilon$ as a function of $N_{\mathrm{shot}}$ for the velocity component $u_y$ under the same conditions as in panel (a).}
\end{figure}

\begin{figure}

\raggedright
(a)\\
\includegraphics[width=0.95\linewidth]{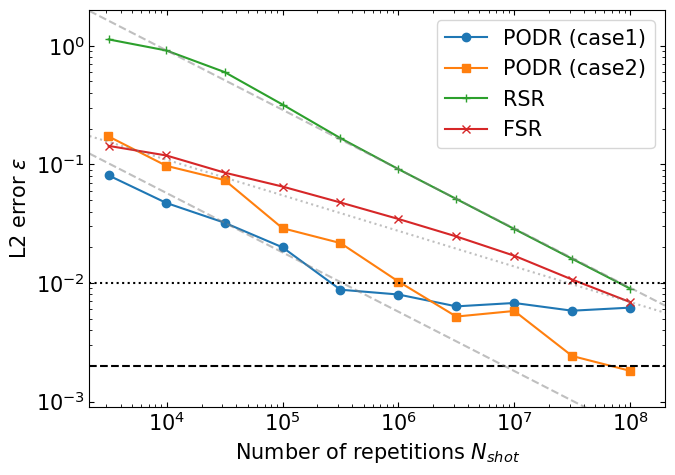}

\vspace{0.5em}

(b)\\
\includegraphics[width=0.95\linewidth]{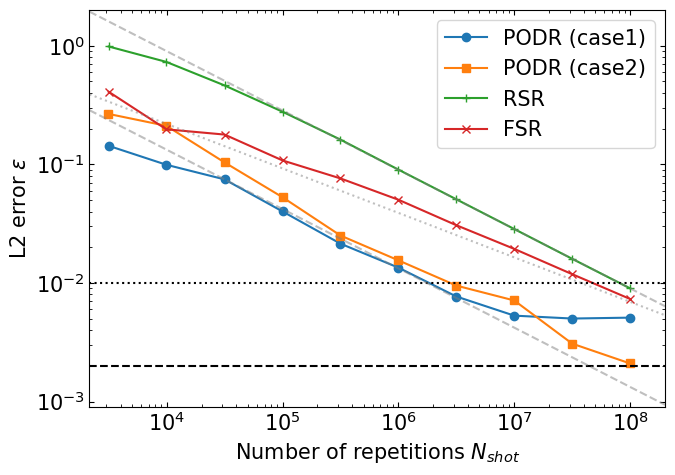}
\caption{\label{fig:nshotuy_kar0}
(a) Readout error $\varepsilon$ as a function of the number of shots for the velocity field $u_x$ of the 2D K\'arm\'an vortex street. The gray dashed lines indicate $\mathcal{O}(N_{\mathrm{shot}}^{-1/2})$ scaling for RSR and PODR. The gray dotted line indicates $\mathcal{O}(N_{\mathrm{shot}}^{-3/10})$ scaling for FSR. Case~1 and Case~2 correspond to the parameter settings listed in Table~\ref{tab:nshot_setting_kar}. The corresponding prescribed errors are indicated by a black dotted horizontal line for Case~1 and a black dashed horizontal line for Case~2. (b) Readout error as a function of $N_{\mathrm{shot}}$ for the velocity component $u_y$ under the same conditions as in panel (a). The gray dotted line indicates the $\mathcal{O}(N_{\mathrm{shot}}^{-3/8})$ scaling for FSR.}
\end{figure}

\subsection{Shot efficiency}
\label{subsec:nshot}
Fig.~\ref{fig:2}(a) shows the dependence of the readout error $\varepsilon$ on $N_{\mathrm{shot}}$ for the velocity component $u_x$ of the 2D lid-driven cavity flow, and Fig.~\ref{fig:2}(b) shows the corresponding result for $u_y$.

We first analyze the readout performance of RSR and FSR. In Fig.~\ref{fig:2}, the gray dotted lines indicating the FSR scaling correspond to the estimates in Appendix D.2 of Ref.~\cite{Huang:2025exw}, where the smoothness parameter is $s = 4/3$. This scaling arises from reduced efficiency due to the nonsmooth behavior at the cavity corners. As discussed in Sec.~\ref{meth:complexity}, the scaling of $N_{\mathrm{shot}}$ with respect to $\varepsilon$ is less favorable for FSR, $\mathcal{O}\!\left((1/\varepsilon)^{2+s}\right)$, than for RSR, $\mathcal{O}\!\left(N(1/\varepsilon)^2\right)$. However, the prefactor associated with the grid size $N$ is smaller for FSR, and FSR therefore achieves lower readout error than RSR in the regime $N_{\mathrm{shot}} \lesssim 10^8$, as shown in Fig.~\ref{fig:2}.

For PODR, the total readout error is bounded by the sum of three contributions,
$E_{\mathrm{proj}}$, $E_{\mathrm{enc}}$, and $E_{\mathrm{sam}}$, as described in Sec.~\ref{readouterror}. Among these terms, $E_{\mathrm{proj}}$ and $E_{\mathrm{enc}}$ are fixed once $n_{\mathrm{b}}$ and $\{ \chi_i\}_{i=1}^{n_{\mathrm{b}}}$ are selected, whereas $E_{\mathrm{sam}}$ decreases as $N_{\mathrm{shot}}^{-1/2}$. Therefore, at small $N_{\mathrm{shot}}$, the sampling error dominates the total error and PODR follows the $\mathcal{O}(N_{\mathrm{shot}}^{-1/2})$ scaling.
As $N_{\mathrm{shot}}$ increases, $E_{\mathrm{sam}}$ becomes smaller than the fixed contribution $E_{\mathrm{proj}}+E_{\mathrm{enc}}$, and the total error saturates. In Fig.~\ref{fig:2}, the crossover point approximately corresponds to the intersection between the $N_{\mathrm{shot}}^{-1/2}$ sampling-error scaling line and the horizontal prescribed error line. In Case~1, this crossover occurs at approximately $N_{\mathrm{shot}}\sim 3\times 10^5$. In Case~2, the stricter choices of $n_{\mathrm{b}}$ and $\{ \chi_i\}_{i=1}^{n_{\mathrm{b}}}$ reduce $E_{\mathrm{proj}}+E_{\mathrm{enc}}$, and the crossover is shifted to
$N_{\mathrm{shot}}\sim 10^7$.

As shown in Appendix~\ref{app:detailcaverr}, the projection error $E_{\mathrm{proj}}$ exhibits an approximately exponential decrease with respect to $n_{\mathrm{b}}$ over the error range relevant to the present benchmark cases. This behavior suggests $n_{\mathrm{b}}=\mathcal{O}(\log_2(1/\varepsilon))$ within this practically relevant error regime. Combining this with Eq.~(\ref{shotscale}), the required number of shots scales as
\begin{eqnarray}
\label{shotscalecav}
N_{\mathrm{shot}} = \mathcal{O}\!\left((1/\varepsilon)^2(\log_2(1/\varepsilon))^2\right).
\end{eqnarray}
These results demonstrate that, for the 2D lid-driven cavity flow, PODR requires fewer shots than both RSR and FSR to achieve the same error tolerance.

Fig.~\ref{fig:nshotuy_kar0}(a) shows the dependence of the readout error $\varepsilon$ on $N_{\mathrm{shot}}$ for the velocity component $u_x$ of the 2D K\'arm\'an vortex street, and Fig.~\ref{fig:nshotuy_kar0}(b) shows the corresponding result for $u_y$.

The 2D K\'arm\'an vortex street benchmark exhibits the same qualitative error behavior as the steady lid-driven cavity flow. As shown in Fig.~\ref{fig:nshotuy_kar0}, PODR follows the $N_{\mathrm{shot}}^{-1/2}$ scaling at small shot numbers, where the sampling error is the dominant contribution, and the error subsequently saturates when the fixed projection and encoding errors become dominant. Fig.~\ref{fig:nshotuy_kar0} indicates that the PODR shot-efficiency advantage is not limited to the steady-flow benchmark. However, compared with the cavity-flow case, the K\'arm\'an vortex street requires a larger number of POD basis functions, especially for $u_y$, reflecting the richer spatial structure of the vortex-shedding dynamics.

\begin{figure}[t]
\centering
\includegraphics[width=1\linewidth]{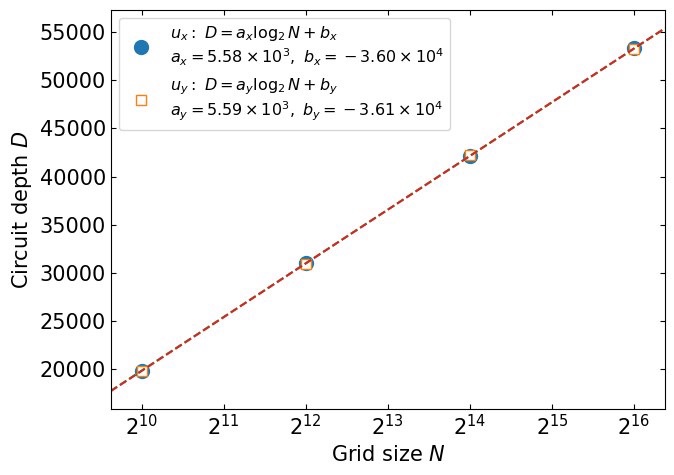}
\caption{\label{fig:uxcount}Circuit depth $D$ for the representative POD basis function of the steady 2D lid-driven cavity flow.
For each grid size $N$, the representative basis function is chosen as the one with the largest bond dimension under the Case~2 settings listed in Table~\ref{tab:nshot_setting}.
The dashed lines show least-squares fits to the data for $u_x$ and $u_y$.}
\end{figure}

\subsection{Circuit depth}
\label{subsec:gate}
Fig.~\ref{fig:uxcount} shows the circuit depth $D$ of the representative POD basis encoding circuit for the 2D lid-driven cavity flow in Case~2.
The representative circuit is chosen as the selected POD basis function with the largest bond dimension, providing an upper estimate of the depth.
After transpilation using Qiskit~\cite{Javadi-Abhari:2024kbf}, $D$ is consistent with logarithmic scaling in $N$ for both $u_x$ and $u_y$, indicating that the encoding step in PODR remains efficient over the tested grid sizes. 

This favorable scaling of $D$ can be attributed to the fact that the dominant POD basis functions typically exhibit simple spatial structures, while the maximum bond dimension satisfies $\chi = \mathcal{O}(\log_2(1/\varepsilon))$ (see Appendix~\ref{app:detailcaverr}). Combining this with Eq.~(\ref{podgatecomp}), the gate complexity of each encoding circuit is given by
\begin{eqnarray}
\label{podgatecompcav}
C_U = \mathcal{O}\!\left(\mathrm{polylog}(N,1/\varepsilon)\right).
\end{eqnarray}
Combining Eqs.~(\ref{shotscalecav}) and (\ref{podgatecompcav}), the overall quantum complexity, defined as the gate complexity multiplied by the number of shots, becomes $\mathcal{O}\!\left((1/\varepsilon)^2 \mathrm{polylog}(N,1/\varepsilon)\right)$. 

The scalability of PODR depends on keeping both the number of selected POD basis functions and their bond dimensions sufficiently small. Stricter readout error tolerances generally require more POD basis functions. Appendix~\ref{basisvis} illustrates this point for the steady 2D lid-driven cavity flow, showing that higher-index POD basis functions exhibit more complex spatial structures and tend to require larger bond dimensions. In addition, more complex solution structures, such as turbulent flows~\cite{Michailidis:2024kcs,Gourianov:2021tot}, are also expected to require larger bond dimensions. These observations highlight an important direction for future optimization to maintain scalability in more challenging physical regimes.

\section{Conclusion}
\label{conc}
In this work, we have proposed a quantum state readout method, termed proper orthogonal decomposition-based readout (PODR). The method follows an offline-online decomposition framework, in which POD basis functions are constructed in the offline stage and the solution is reconstructed from corresponding coefficients in the online stage. By leveraging the orthogonality and problem-specific nature of the POD basis functions, PODR improves readout efficiency compared with conventional methods.

We analytically estimated the computational resources required for PODR in both the offline and online stages. Our analysis shows that the cost of the online stage depends on the number of POD basis functions and their bond dimensions in the matrix product state (MPS) representation, while it exhibits only a polylogarithmic dependence on the grid size. Numerical evaluations using computational fluid dynamics (CFD) benchmark problems indicated that both the number of basis functions and the bond dimensions scale logarithmically with respect to the error tolerance for the benchmark problems considered. Furthermore, PODR reduces the online computational cost compared with conventional methods, including Fourier space readout (FSR) and real-space readout (RSR).

These results indicate that PODR is particularly well suited to practical simulations in computer-aided engineering (CAE) and CFD, where repeated computations are required for varying physical parameters. In such settings, the offline stage needs to be performed only once, and the resulting POD basis functions can be reused for efficient readout in multiple online evaluations.

Finally, PODR may be applicable beyond CAE/CFD to other areas in which readout constitutes a computational bottleneck, such as quantum financial simulations and quantum image processing. Exploring these applications is an important direction for future work.

\section{Acknowledgments}
We thank members of Quemix Inc. and SCSK Corporation for valuable discussions and insights related to this work.

\appendix

\begin{figure}[t]

\raggedright
(a)\\
\includegraphics[width=0.95\linewidth]{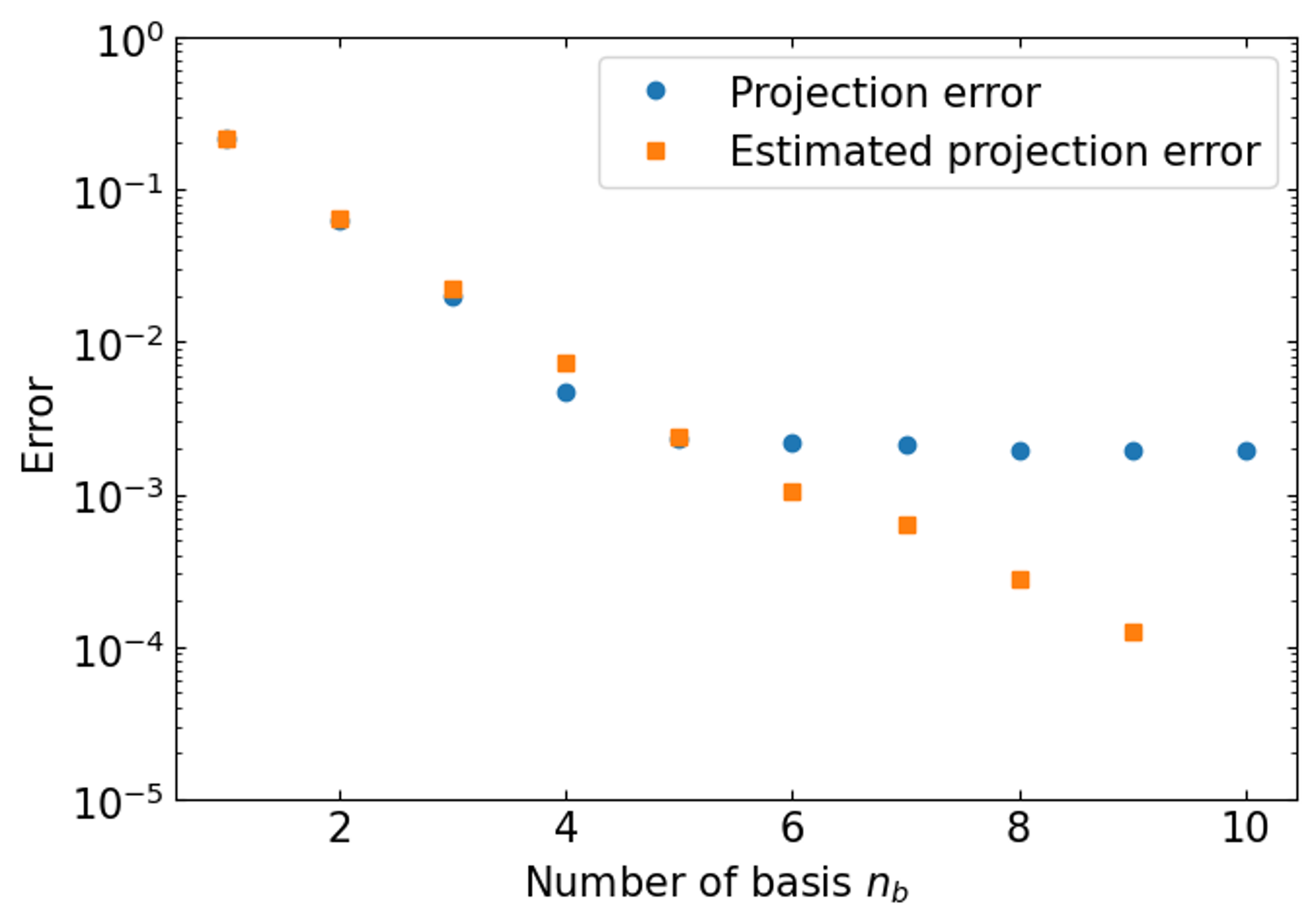}

\vspace{0.5em}

(b)\\
\includegraphics[width=0.95\linewidth]{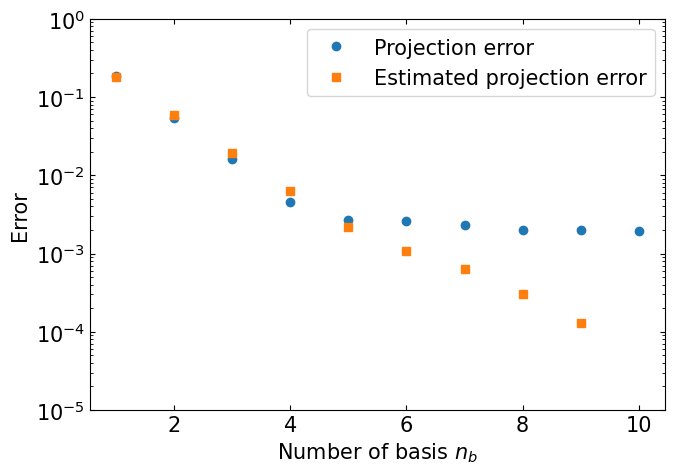}
\caption{\label{fig:projestuy}(a) Projection error and the corresponding estimated projection error at $Re = 950$ as functions of $n_{\mathrm{b}}$ for the velocity component $u_x$. (b) Corresponding results for the velocity component $u_y$.}
\end{figure}

\begin{figure*}[t]
\centering
\includegraphics[width=0.95\linewidth]{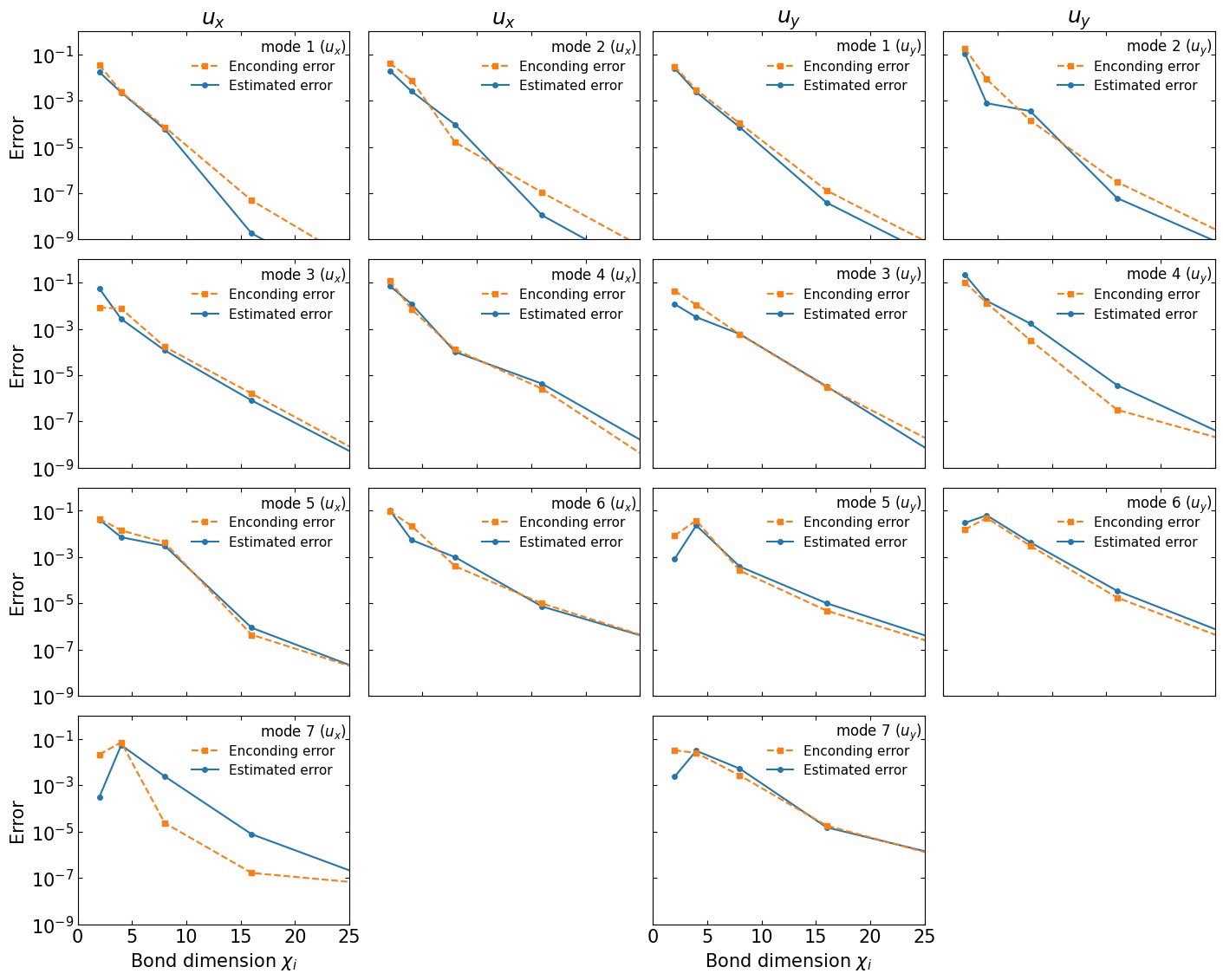}
\caption{\label{fig:chi}Encoding error of each POD basis function at $Re = 950$, $\left| \langle \boldsymbol{x} \mid \boldsymbol{\delta}_i \rangle \right|$ (dashed lines), and the corresponding per-basis terms in the estimated encoding
error (solid lines) as functions of the maximum bond dimension $\chi_i$.}
\end{figure*}

\section{Derivation of the estimated errors}
\label{estdousitu}
We derive the estimated projection and encoding errors introduced in
Sec.~\ref{shuhou}. These quantities approximate the corresponding error terms in Eq.~(\ref{leqq}) using information available before readout. Because the target solution $\boldsymbol{x}$ is not available, the exact values of $E_{\mathrm{proj}}$ and $E_{\mathrm{enc}}$ cannot be computed in the offline stage.

We first introduce the mean-squared coefficient of each POD basis function over the
snapshot ensemble. Let the snapshot matrix be
\begin{equation}
\boldsymbol{S}=
\bigl[
\boldsymbol{x}_1\;\; \boldsymbol{x}_2\;\; \cdots\;\; \boldsymbol{x}_M
\bigr],
\end{equation}
and let its singular value decomposition be
\begin{equation}
\label{app1}
\boldsymbol{S} = \boldsymbol{U} \boldsymbol{\Sigma} \boldsymbol{V}^{\top}.
\end{equation}
Here, $\boldsymbol{U}$ and $\boldsymbol{V}$ contain the left and right singular vectors, respectively, and $\boldsymbol{\Sigma}=\mathrm{diag}(\sigma_1,\ldots,\sigma_M)$ is the diagonal matrix of singular values.
The columns of $\boldsymbol{U}=[\boldsymbol{u}_1,\ldots,\boldsymbol{u}_M]$ are the POD basis functions. For the $j$th snapshot $\boldsymbol{x}_j$, the coefficient associated with the $i$th POD basis function is given by
\begin{equation}
a_i = \boldsymbol{u}_i^{\top} \boldsymbol{x}_j.
\end{equation}
The mean squared value of this coefficient over all snapshots is
\begin{equation}
\label{2zixyouheikinn}
\mathbb{E}\!\left[|a_i|^2\right]
=
\frac{1}{M}\sum_{j=1}^{M}(\boldsymbol{u}_i^{\top} \boldsymbol{x}_j)^2
=
\frac{1}{M}\|\boldsymbol{u}_i^{\top}\boldsymbol{S}\|_{2}^2.
\end{equation}
Using Eq.~(\ref{app1}), we obtain
\begin{align}
\boldsymbol{u}_i^{\top} \boldsymbol{S}
=
\boldsymbol{u}_i^{\top} \boldsymbol{U} \boldsymbol{\Sigma} \boldsymbol{V}^{\top}
=
\sigma_i \boldsymbol{v}_i^{\top}.
\end{align}
Since $\boldsymbol{v}_i$ is normalized,
\begin{equation}
\left\|
\boldsymbol{u}_i^{\top} \boldsymbol{S}
\right\|_2^2
=
\sigma_i^2
\left\|
\boldsymbol{v}_i^{\top}
\right\|_2^2
=
\sigma_i^2.
\end{equation}
Therefore, from Eq.~(\ref{2zixyouheikinn}), we obtain
\begin{equation}
\label{rms}
\mathbb{E}\!\left[|a_i|^2\right]
=
\frac{\sigma_i^2}{M}.
\end{equation}
This completes the evaluation of the mean-squared coefficient
of each POD basis function over the snapshot ensemble.

To construct the estimated errors, we assume that the target
solution is well represented by the snapshot ensemble.
Since the POD coefficients of the target solution are not
available in the offline stage, we use the root-mean-square
(RMS) magnitude of each POD coefficient over the snapshot
ensemble as a representative magnitude of the corresponding
unknown target coefficient. From Eq.~(\ref{rms}), this RMS magnitude
is given by
\begin{equation}
c_i^{\mathrm{rms}}
\equiv
\sqrt{\mathbb{E}[|a_i|^2]}
=
\frac{\sigma_i}{\sqrt{M}}.
\label{eq:rms_coefficient}
\end{equation}
Here, $c_i^{\mathrm{rms}}$ represents a characteristic magnitude inferred from the snapshot ensemble, rather than the exact coefficient of an individual target solution. Using these representative magnitudes, the projection and encoding errors are estimated from information available in the offline stage.
\begin{align}
\label{errestdousitu}
E_{\mathrm{proj}}
&=
\left\|
\sum_{i=n_{\mathrm{b}}+1}^{N}
c_i \boldsymbol{u}_i
\right\|_2
\nonumber\\
&\approx
\left\|
\sum_{i=n_{\mathrm{b}}+1}^{M}
c_i^{\mathrm{rms}}
\boldsymbol{u}_i
\right\|_2
\nonumber\\
&=
\left\|
\sum_{i=n_{\mathrm{b}}+1}^{M}
\sqrt{\frac{\sigma_i^2}{M}}
\boldsymbol{u}_i
\right\|_2 .
\end{align}
In the second line of Eq.~(\ref{errestdousitu}), we used the fact that $M<N$ and that $\sigma_i=0$ for $i>M$, since the snapshot matrix has at most $M$ nonzero singular values. Therefore, the estimated projection error can be written as
\begin{equation}
\label{projest1}
E^{\mathrm{est}}_{\mathrm{proj}} = 
\sqrt{\frac{1}{M} \sum_{i=n_{\mathrm{b}}+1}^{M} \sigma_i^2}.
\end{equation}
Fig.~\ref{fig:projestuy} compares $E_{\mathrm{proj}}$ with $E^{\mathrm{est}}_{\mathrm{proj}}$ as functions of $n_{\mathrm{b}}$ for the velocity fields $u_x$ and $u_y$ in the steady 2D lid-driven cavity flow at $Re=950$ on a $256^2$ grid. The estimated error captures the initial decrease of the projection error for small $n_{\mathrm{b}}$, but it underestimates the true projection error when higher-order modes are included. In particular, the true projection error saturates at approximately $2\times 10^{-3}$ for this target state, whereas $E^{\mathrm{est}}_{\mathrm{proj}}$ continues to decrease. This discrepancy indicates that $E^{\mathrm{est}}_{\mathrm{proj}}$ should be interpreted as a snapshot-statistical criterion, rather than as a guaranteed prediction of the projection error for an individual target state.

Next, we construct an estimate of the encoding error in
Eq.~(\ref{leqq}) using the RMS coefficient magnitudes
$c_i^{\mathrm{rms}}$:
\begin{align}
E_{\mathrm{enc}}
&=
\left\|
\sum_{i=1}^{n_{\mathrm{b}}}
\left(
c_i-\langle \boldsymbol{x}|
\tilde{\boldsymbol{u}}_i\rangle
\right)
\boldsymbol{u}_i
\right\|_2
\nonumber\\
&\approx
\left\|
\sum_{i=1}^{n_{\mathrm{b}}}
\left(
c_i^{\mathrm{rms}}
-
\sum_{j=1}^{n_{\mathrm{b}}}
c_j^{\mathrm{rms}}
\langle
\boldsymbol{u}_j|
\tilde{\boldsymbol{u}}_i
\rangle
\right)
\boldsymbol{u}_i
\right\|_2
\nonumber\\
&=
\left\|
\sum_{i=1}^{n_{\mathrm{b}}}
\left(
\sqrt{\frac{\sigma_i^2}{M}}
-
\sum_{j=1}^{n_{\mathrm{b}}}
\sqrt{\frac{\sigma_j^2}{M}}
\langle
\boldsymbol{u}_j|
\tilde{\boldsymbol{u}}_i
\rangle
\right)
\boldsymbol{u}_i
\right\|_2 .
\end{align}
In going from the first to the second line, we use only the first $n_{\mathrm{b}}$ MPS approximated POD basis functions, because the reconstruction is truncated to the selected POD subspace. Therefore, $E^{\mathrm{est}}_{\mathrm{enc}}$ can be written as
\begin{equation}
\label{encest1}
E^{\mathrm{est}}_{\mathrm{enc}}
=\sqrt{\sum_{i=1}^{n_{\mathrm{b}}} \left|
 \sqrt{\frac{\sigma_i^{2}}{M}}- \sum_{j=1}^{n_{\mathrm{b}}} \sqrt{\frac{\sigma_j^{2}}{M}}\,\langle \tilde{\boldsymbol{u}}_i \mid \boldsymbol{u}_j \rangle\right|^2}. 
\end{equation}
As with $E_{\mathrm{proj}}$, this approximation should be interpreted as a snapshot-statistical estimate rather than an exact prediction for an individual target solution. Fig.~\ref{fig:chi} plots the encoding error for each selected POD basis function $\left| \langle \boldsymbol{x} \mid \boldsymbol{\delta}_i \rangle \right|$, and the corresponding per-basis terms in the estimated encoding error as functions of the maximum bond dimension $\chi_{i}$ for the velocity fields $u_x$ and $u_y$ in the steady 2D lid-driven cavity flow setting of Sec.~\ref{setcav} on a $256^2$ grid. As shown in Fig.~\ref{fig:chi}, the estimated error shows reasonable agreement with the actual encoding error.

\section{Details of MPS approximation of POD basis functions}
\label{app:mps}

In the main text, the offline stage of PODR includes the construction of MPS approximations to the selected POD basis functions. This appendix provides the details of this MPS approximation procedure. In particular, we describe the ordering of the MPS cores, the selection of the maximum bond dimension, and the associated computational cost.

We first specify how the two-dimensional POD basis functions are mapped to MPS. For a two-dimensional computational domain discretized on a spatial grid, each coordinate is represented in binary form. Specifically, the spatial coordinates $x$ and $y$ are encoded using
$n_x$ and $n_y$ qubits, respectively, as
\begin{equation}
x
=
\sum_{k=1}^{n_x} x_k 2^{k-1},
\qquad
y
=
\sum_{l=1}^{n_y} y_l 2^{l-1},
\end{equation}
where $x_k, y_l \in \{0,1\}$ denote the binary digits. The physical indices of the MPS are ordered as
\begin{equation}
(x_1,x_2,\ldots,x_{n_x},
y_1,y_2,\ldots,y_{n_y}),
\end{equation}
which determines the sequence of tensor cores in the MPS decomposition.

Once the core ordering is fixed, each POD basis function
$\boldsymbol{u}_i\in\mathbb{R}^{N}$ is approximated by an MPS. The approximation accuracy is controlled by the maximum bond dimension $\chi_i$: increasing $\chi_i$ generally improves the approximation accuracy, but also increases the computational cost. Therefore, for each POD basis function, we select $\chi_i$ that satisfies the prescribed threshold for the estimated encoding error $E_{\mathrm{enc}}^{\mathrm{est}}$.

We next describe the procedure used to select $\chi_i$ by searching over candidate maximum bond dimensions. From the viewpoint of quantum circuit implementation, a bond dimension $\chi_i$ must be embedded into a Hilbert space whose dimension is a power of two~\cite{PRXQuantum.2.010342,SciPostPhys.14.5.117}. Hence, all values of $\chi_i$ within
\begin{equation}
2^{m-1}<\chi_i\le 2^m
\end{equation}
lead to the same circuit size. It is thus sufficient to consider the candidate set
\begin{equation}
\{1,2,4,8,\ldots\}.
\end{equation}
Therefore, searching up to $\chi\equiv\max_{1\le i\le n_{\mathrm{b}}}\chi_i$ requires only $\mathcal{O}(\log_2\chi)$ candidates.

Finally, we estimate the computational cost of this procedure. Constructing an MPS approximation of a vector of dimension $N$ with maximum bond dimension $\chi$ requires
\begin{equation}
\mathcal{O}(N\chi^2),
\end{equation}
operations. Since the MPS approximation is evaluated for
$\mathcal{O}(n_\mathrm{b}\log_2\chi)$ candidate bond dimensions, the cost of determining the bond dimension for POD basis functions is
\begin{equation}
\mathcal{O}(Nn_\mathrm{b}\chi^2\log_2\chi).
\end{equation}

\section{Error-bound inequality}
\label{app:simeq}
In Sec.~\ref{readouterror}, we introduced the error bound in Eq.~(\ref{leqq}). In this section, we provide its derivation.

We first expand the solution state $\boldsymbol{x}$ in a complete orthonormal basis that contains the POD basis functions. Let the snapshot matrix $\boldsymbol{S} \in \mathbb{R}^{N \times M}$ be decomposed as $\boldsymbol{S} = \boldsymbol{U}\boldsymbol{\Sigma}\boldsymbol{V}^{\top}$, and let $\{\boldsymbol{u}_i\}_{i=1}^{M}$ denote the column vectors of $\boldsymbol{U}$, i.e., the POD basis functions. A complete orthonormal basis $\{\boldsymbol{u}_i\}_{i=1}^{N}$ is obtained by augmenting these vectors with an orthonormal complement $\{\boldsymbol{u}_i\}_{i=M+1}^{N}$. Using this basis, the unknown solution state is written as
\begin{equation}
\label{seikai}
\boldsymbol{x}
=
\sum_{i=1}^{N} c_i \, \boldsymbol{u}_i,
\qquad
\sum_{i=1}^{N} c_i^2 = 1,
\end{equation}
where $c_i \in \mathbb{R}$ are the coefficients.
As described in Sec.~\ref{sub:online}, the reconstructed state $\tilde{\boldsymbol{x}}$ is given by
\begin{equation}
\label{saikouseiseikai}
\tilde{\boldsymbol{x}}
=
\sum_{i=1}^{n_{\mathrm{b}}} \tilde{c}_i \boldsymbol{u}_i.
\end{equation}

From Eqs.~(\ref{seikai}) and (\ref{saikouseiseikai}), the $L_2$ error between $\boldsymbol{x}$ and $\tilde{\boldsymbol{x}}$ is
\begin{widetext}
\begin{align}
\label{ineq}
\varepsilon
&=
\left\lVert
\sum_{i=1}^{N} c_i \boldsymbol{u}_i
-
\sum_{i=1}^{n_{\mathrm{b}}} \tilde{c}_i \boldsymbol{u}_i
\right\rVert_{2}
\\ \nonumber
&=
\left\lVert
\sum_{i=1}^{n_{\mathrm{b}}} c_i \boldsymbol{u}_i
-
\sum_{i=1}^{n_{\mathrm{b}}} \langle \boldsymbol{x} \mid \tilde{\boldsymbol{u}}_i \rangle \boldsymbol{u}_i
+
\sum_{i=1}^{n_{\mathrm{b}}} \langle \boldsymbol{x} \mid \tilde{\boldsymbol{u}}_i \rangle \boldsymbol{u}_i
-
\sum_{i=1}^{n_{\mathrm{b}}} \tilde{c}_i \boldsymbol{u}_i
+
\sum_{i=n_{\mathrm{b}}+1}^{N} c_i \boldsymbol{u}_i
\right\rVert_{2}
\\ \nonumber
&\le
\left\lVert
\sum_{i=1}^{n_{\mathrm{b}}} \bigl(c_i - \langle \boldsymbol{x} \mid \tilde{\boldsymbol{u}}_i \rangle\bigr) \boldsymbol{u}_i
\right\rVert_{2}
+
\left\lVert
\sum_{i=1}^{n_{\mathrm{b}}} \bigl(\tilde{c}_i - \langle \boldsymbol{x} \mid \tilde{\boldsymbol{u}}_i \rangle\bigr) \boldsymbol{u}_i
\right\rVert_{2}
+
\left\lVert
\sum_{i=n_{\mathrm{b}}+1}^{N} c_i \boldsymbol{u}_i
\right\rVert_{2}.
\end{align}
\end{widetext}

We next bound the three terms in Eq.~(\ref{ineq}). For the first term, using $c_i = \langle \boldsymbol{x} \mid \boldsymbol{u}_i \rangle$, we obtain
\begin{align}
\label{enc1}
E_{\mathrm{enc}}
&=
\left\lVert
\sum_{i=1}^{n_{\mathrm{b}}}
\bigl(c_i - \langle \boldsymbol{x} \mid \tilde{\boldsymbol{u}}_i \rangle\bigr)\boldsymbol{u}_i
\right\rVert_{2}
\nonumber \\
&=
\sqrt{\sum_{i=1}^{n_{\mathrm{b}}}
\left\lvert \langle \boldsymbol{x} \mid \boldsymbol{\delta}_i \rangle \right\rvert^2},
\end{align}
where $\ket{\boldsymbol{\delta}_i}
= \ket{\boldsymbol{u}_i} - \ket{\tilde{\boldsymbol{u}}_i}$.

To bound the second term in Eq.~(\ref{ineq}), we evaluate the sampling error of $\tilde{c}_i$.
We assume that the total number of shots is distributed equally among the selected POD basis functions, and denote the number of shots assigned to each basis function by $N_{\mathrm{shot}}^{\mathrm{basis}}$.
As described in Sec.~\ref{sub:online}, the coefficient $\tilde{c}_i$ is given by
$\tilde{c}_i = 2\tilde{p}_{0,i} - 1$, where
$\tilde{p}_{0,i} = Z_{0,i}/N_{\mathrm{shot}}^{\mathrm{basis}}$.
Here, $Z_{0,i}$ denotes the number of times $\ket{0}$ is observed in the Hadamard test for the $i$th POD basis function.
Thus, $Z_{0,i}$ follows the binomial distribution
$\mathrm{B}(N_{\mathrm{shot}}^{\mathrm{basis}}, p_{0,i})$, where $p_{0,i}$ is the probability of obtaining $\ket{0}$.
Therefore, $\tilde{c}_i$ is an unbiased estimator of $\langle \boldsymbol{x} \mid \tilde{\boldsymbol{u}}_i \rangle$, with
\begin{equation}
\mathbb{E}[\tilde{c}_i] = 2p_{0,i} - 1,
\qquad
\mathrm{Var}[\tilde{c}_i] = \frac{4p_{0,i}(1-p_{0,i})}{N_{\mathrm{shot}}^{\mathrm{basis}}}.
\end{equation}
By Chebyshev's inequality, for any $\beta > 0$,
\begin{align}
\label{chev}
\mathbb{P}\!\left(
\left| \tilde{c}_i - \langle \boldsymbol{x} \mid \tilde{\boldsymbol{u}}_i \rangle \right|
\ge
\beta \sqrt{\frac{4p_{0,i}(1-p_{0,i})}{N_{\mathrm{shot}}^{\mathrm{basis}}}}
\right)
&\le
\frac{1}{\beta^2}.
\end{align}
Thus, for each coefficient, Chebyshev's inequality gives the bound
\begin{equation}
\left| \tilde{c}_i - \langle \boldsymbol{x} \mid \tilde{\boldsymbol{u}}_i \rangle \right|
\le
\beta \sqrt{\frac{4p_{0,i}(1-p_{0,i})}{N_{\mathrm{shot}}^{\mathrm{basis}}}}.
\end{equation}
Using this coefficient-wise bound, the sampling contribution is bounded as
\begin{align}
E_{\mathrm{sam}}
&=
\left\lVert
\sum_{i=1}^{n_{\mathrm{b}}}
\bigl(\tilde{c}_i - \langle \boldsymbol{x} \mid \tilde{\boldsymbol{u}}_i \rangle\bigr)\boldsymbol{u}_i
\right\rVert_{2}
\nonumber \\
&\leq
\beta \sqrt{
\sum_{i=1}^{n_{\mathrm{b}}}
\frac{4p_{0,i}(1 - p_{0,i})}{N_{\mathrm{shot}}^{\mathrm{basis}}}
}
\nonumber \\
&\le
\beta \sqrt{\frac{n_{\mathrm{b}}}{N_{\mathrm{shot}}^{\mathrm{basis}}}}.
\end{align}

The third term corresponds to the component of $\boldsymbol{x}$ that cannot be represented by the first $n_{\mathrm{b}}$ POD basis functions and therefore defines the projection error:
\begin{align}
\label{prog1}
E_{\mathrm{proj}}
&=
\left\lVert
\sum_{i=n_{\mathrm{b}}+1}^{N} c_i \boldsymbol{u}_i
\right\rVert_{2}
\nonumber \\
&=
\| \boldsymbol{x} - \boldsymbol{U}_{n_{\mathrm{b}}} \boldsymbol{U}_{n_{\mathrm{b}}}^{\top} \boldsymbol{x} \|_{2}.
\end{align}
This completes the derivation of Eq.~(\ref{leqq}).

\begin{figure}[t]

\raggedright
(a)\\
\includegraphics[width=0.95\linewidth]{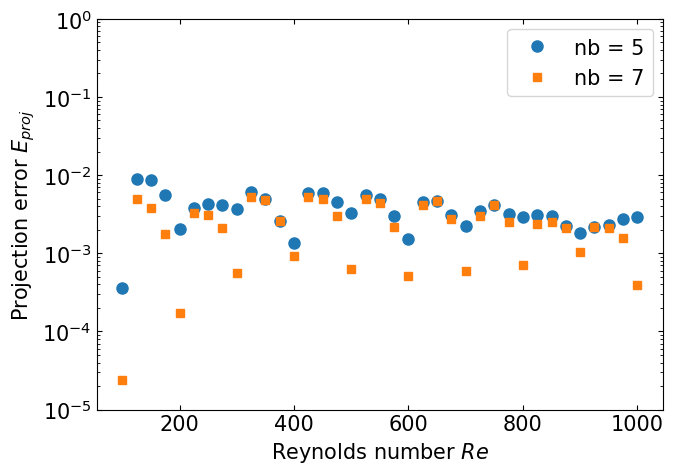}

\vspace{0.5em}

(b)\\
\includegraphics[width=0.95\linewidth]{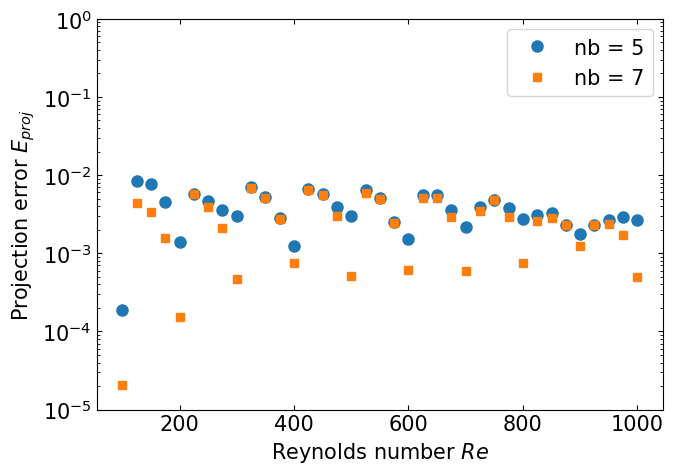}

\caption{\label{fig:redip}(a) The projection errors for Case~1 and Case~2 at each Reynolds number $Re$. The Reynolds numbers $Re=100k\ (k=1,2,\ldots,10)$ are included in the snapshot matrix, whereas the other $Re$ are not.
(b) Corresponding results for the velocity component $u_y$ of the steady 2D lid-driven cavity flow.}
\end{figure}

\begin{figure}

\raggedright
(a)\\
\includegraphics[width=0.95\linewidth]{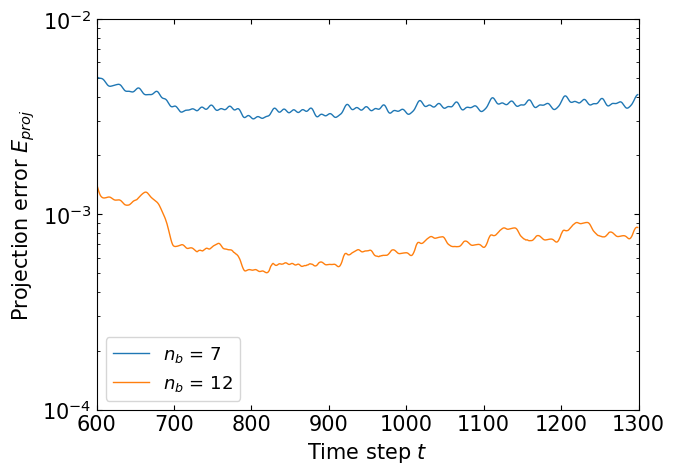}

\vspace{0.5em}

(b)\\
\includegraphics[width=0.95\linewidth]{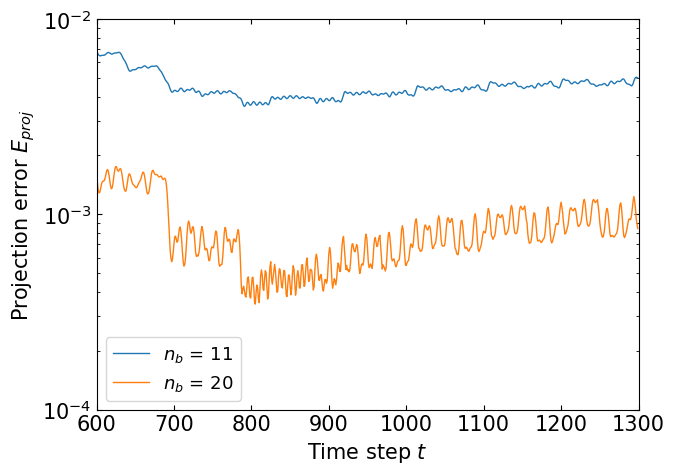}
\caption{\label{fig:nshotuykar}(a) The projection errors for Case~1 and Case~2 at each time step $t$ for the velocity field $u_x$ of the 2D K\'arm\'an vortex street. Time steps 600 to 1200 are included in the snapshot matrix, whereas time steps 1201 to 1300 are not included. (b) Corresponding results for the velocity component $u_y$ of the 2D K\'arm\'an vortex street.}
\end{figure}

\section{Effect of the snapshot matrix}
\label{app:snapshot_effect}
In this section, we provide the rationale for the choice of the target solution state described in Secs.~\ref{setcav} and \ref{setkar} by examining the effect of the snapshot matrix on the projection error.

In Sec.~\ref{setcav}, we presented the representative target state for $Re=950$. Here, we verify that the projection error remains comparable for the other Reynolds numbers. Fig.~\ref{fig:redip} plots $E_{\mathrm{proj}}$ for the velocity fields $u_x$ and $u_y$ of the steady 2D lid-driven cavity flow on a $256^2$ grid as functions of $Re$, under Case~1 and Case~2 settings defined in Table~\ref{tab:nshot_setting}. Here, $n_{\mathrm{b}}=5$ corresponds to Case~1 and $n_{\mathrm{b}}=7$ corresponds to Case~2. In Fig.~\ref{fig:redip}, we also plot $E_{\mathrm{proj}}$ for $Re=100k$ ($k=1,2,\ldots,10$); these $Re$ are included in the snapshot matrix, and therefore yield smaller values of $E_{\mathrm{proj}}$ than other $Re$. Although $E_{\mathrm{proj}}$ varies with $Re$, it remains within a similar range to that at $Re=950$ even without the cases $Re=100k$ ($k=1,2,\ldots,10$). This suggests that the solution can be reconstructed with reasonably consistent error tolerance across these $Re$ values.

In Sec.~\ref{setkar}, we presented the representative target state for $t=1273$. Here, we verify that the projection error remains comparable for the other time steps.
Fig.~\ref{fig:nshotuykar} shows $E_{\mathrm{proj}}$ for Case~1 and Case~2 at each time step for the velocity fields $u_x$ and $u_y$ of the 2D K\'arm\'an vortex street, with the parameter settings given in Table~\ref{tab:nshot_setting_kar}. As seen in Fig.~\ref{fig:nshotuykar}, PODR can be applied to all target time steps from $t=1201$ to $1300$, in the same manner as for $t=1273$.

\section{Details of stream function}
\label{streamf}
In Sec.~\ref{meth:vis}, we introduce the visualization setting used for PODR, and here we provide additional details on the stream function defined in that section.

The problems considered in Sec.~\ref{meth:nshot} involve incompressible flows, for which a stream function can be defined. The stream function $\psi(x,y)$ satisfies
\begin{equation}
u_x(x,y)=\frac{\partial \psi}{\partial y}(x,y),
\qquad
u_y(x,y)=-\frac{\partial \psi}{\partial x}(x,y).
\end{equation}
In this study, $\psi(x,y)$ is computed by integrating the horizontal
velocity component in the vertical direction from the lower boundary:
\begin{equation}
\psi(x,y)
=
\int_{y_{\min}}^{y} u_x(x,\eta)\,\mathrm{d}\eta
+
C(x),
\end{equation}
where $C(x)$ is an integration constant that may depend on $x$.

To determine the remaining freedom in $C(x)$, we impose consistency with
$u_y=-\partial_x\psi$. Differentiating the above expression with respect
to $x$, we obtain
\begin{align}
-\frac{\partial \psi}{\partial x}(x,y)
&=
-\int_{y_{\min}}^{y}
\frac{\partial u_x}{\partial x}(x,\eta)\,\mathrm{d}\eta
-
\frac{\partial C(x)}{\partial x}.
\end{align}
Using the incompressibility condition
\begin{equation}
\frac{\partial u_x}{\partial x}
+
\frac{\partial u_y}{\partial y}
=0,
\end{equation}
we have
\begin{align}
-\frac{\partial \psi}{\partial x}(x,y)
&=
\int_{y_{\min}}^{y}
\frac{\partial u_y}{\partial \eta}(x,\eta)\,\mathrm{d}\eta
-
\frac{\partial C(x)}{\partial x} \\
&=
u_y(x,y)-u_y(x,y_{\min})-\frac{\partial C(x)}{\partial x}.
\end{align}
Since \(u_y=-\partial_x\psi\), it follows that
\begin{equation}
\frac{\partial C(x)}{\partial x}=-u_y(x,y_{\min}).
\end{equation}
In the present examples, the normal velocity at the lower boundary vanishes,
\begin{equation}
u_y(x,y_{\min})=0.
\end{equation}
Therefore,
\begin{equation}
\frac{\partial C(x)}{\partial x}=0,
\qquad
C(x)=C_0,
\end{equation}
where $C_0$ is a spatially uniform constant.

Since $C_0$ only shifts the reference level of the stream function, we set
$C_0=0$ without loss of generality and use
\begin{equation}
\psi(x,y)
=
\int_{y_{\min}}^{y} u_x(x,\eta)\,\mathrm{d}\eta .
\end{equation}

\begin{figure*}[t]
\centering
\includegraphics[width=0.95\linewidth]{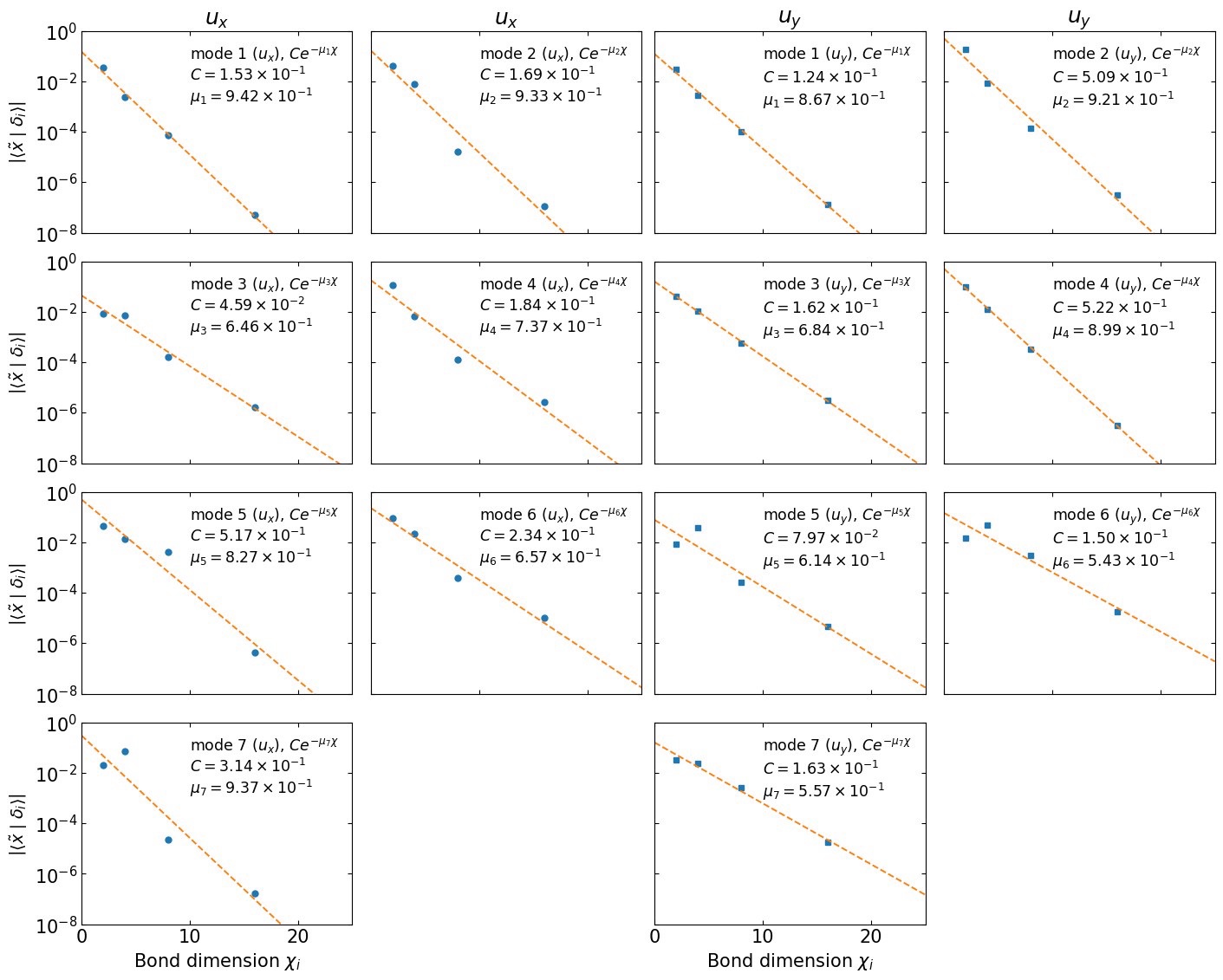}
\caption{\label{fig:uxchi} Dependence of the encoding error of each POD basis function, $\left| \langle \boldsymbol{x} \mid \boldsymbol{\delta}_i \rangle \right|$, on the maximum bond dimension for the target state at $Re = 950$. The dashed lines indicate the least-squares fits to the $u_x$ and $u_y$ data.}
\end{figure*}
\begin{figure}
\centering
\includegraphics[width=0.95\linewidth]{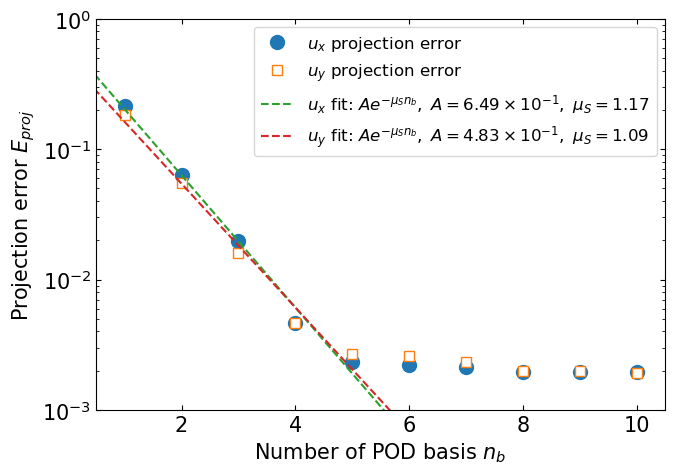}
\caption{\label{fig:uxprojerror} Dependence of the projection error on the number of POD basis functions $n_{\mathrm{b}}$ for the target state at $Re=950$. The dashed lines indicate least-squares exponential fits obtained using the data up to $n_{\mathrm{b}}=5$, where the projection error exhibits an approximately exponential decrease. For larger $n_{\mathrm{b}}$, the target-specific projection error approaches a saturation floor.}
\end{figure}

\begin{figure*}[t]
\includegraphics[width=0.95\linewidth]{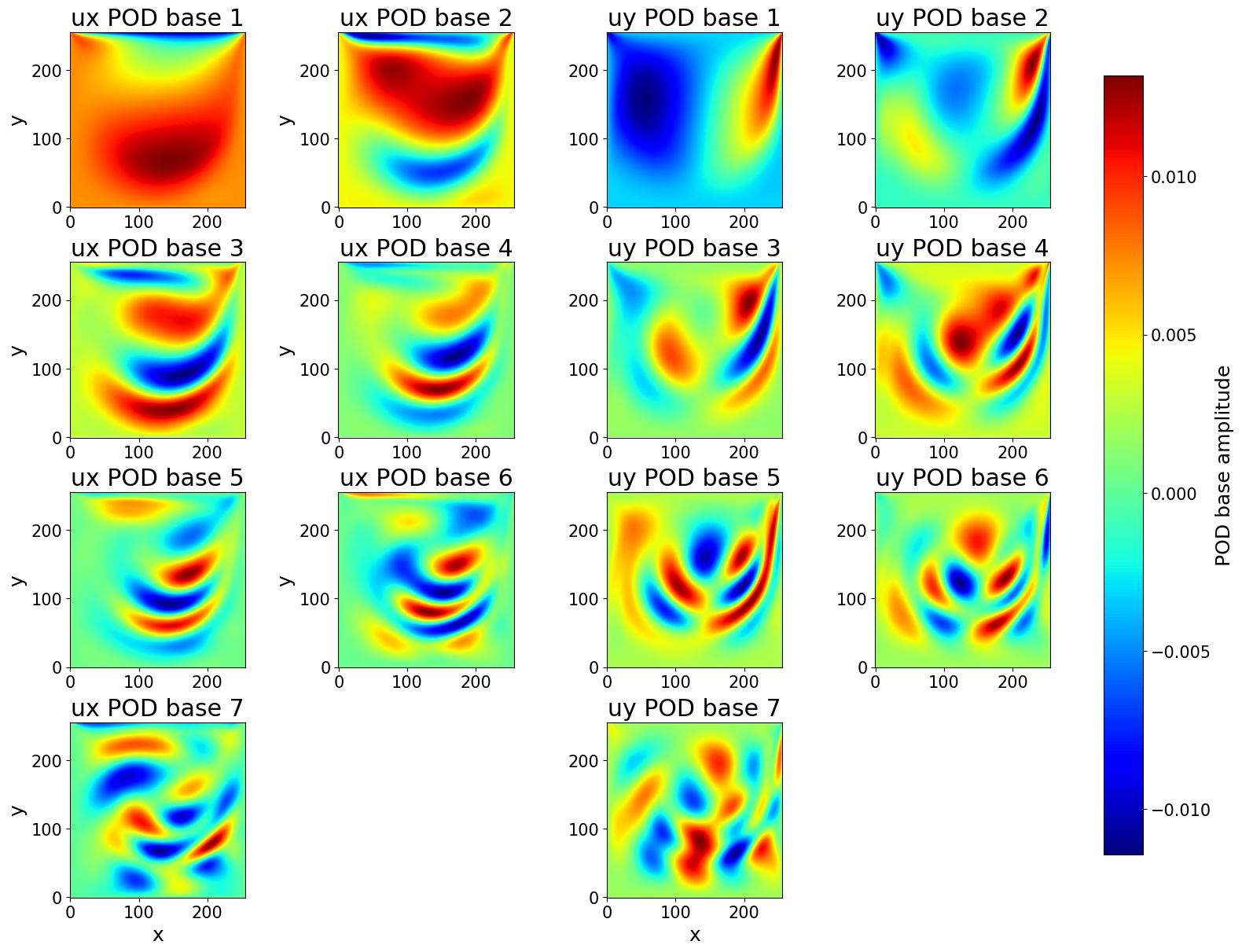}
\caption{\label{fig:basisvis}Two-dimensional color plots of POD basis functions for the velocity fields $u_x$ and $u_y$ of the steady 2D lid-driven cavity flow.}
\end{figure*}

\section{Visualization of POD basis functions}
\label{basisvis}
In this section, we visualize the POD basis functions mentioned in
Sec.~\ref{subsec:gate} to illustrate their spatial structures.

Fig.~\ref{fig:basisvis} shows the two-dimensional color plots of POD basis functions computed for the steady 2D lid-driven cavity flow described in Sec.~\ref{setcav}. The displayed POD basis functions are shown up to $n_{\mathrm{b}}=7$, following the setting of Case~2 listed in Table~\ref{tab:nshot_setting}, and no MPS approximation is applied. 

Fig.~\ref{fig:basisvis} shows that the spatial structures become increasingly complex as the index of the POD basis function increases. 
In the present setting, the highest-index POD basis function among the selected basis functions exhibits the finest spatial structure and requires the largest bond dimension.

\section{Details of the readout error for the steady 2D lid-driven cavity flow}
\label{app:detailcaverr}

As mentioned in Secs.~\ref{subsec:nshot} and \ref{subsec:gate}, we derive the scaling of $N_{\mathrm{shot}}^{\mathrm{basis}}$, $n_{\mathrm{b}}$, and $\chi\equiv\max_{1\le i\le n_{\mathrm{b}}}\chi_i$ with respect to the error tolerance $\varepsilon$, using numerical results for the problem setting described in Sec.~\ref{setcav}.

We first evaluate $E_{\mathrm{enc}}$ in terms of $\{\chi_i\}_{i=1}^{n_{\mathrm{b}}}$. The dominant POD basis functions typically exhibit global and smooth structures, and the required bond dimensions therefore tend to be small (see Appendix~\ref{basisvis} for the POD basis functions). More specifically, Fig.~\ref{fig:uxchi} shows the dependence of $\left| \langle \boldsymbol{x} \mid \boldsymbol{\delta}_i \rangle \right|$ on $\chi_i$ for the steady 2D lid-driven cavity flow on a $256^2$ grid described in Sec.~\ref{setcav}. Over the range of bond dimensions considered here, the numerical results are empirically described by the exponential model

\begin{align}
\label{kateipod}
\left\lvert \langle \boldsymbol{x} \mid \boldsymbol{\delta}_i \rangle \right\rvert
&\le
C_i \exp(-\mu_i \chi_i).
\end{align}
Here, $C_i$ and $\mu_i$ are constants associated with the $i$th POD basis function.
\begin{align}
E_{\mathrm{enc}}
&\le
\sqrt{\sum_{i=1}^{n_{\mathrm{b}}}
\left| C_i \exp(-\mu_i \chi_i) \right|^2}.
\end{align}

We next evaluate $E_{\mathrm{proj}}$ in terms of the number of POD basis functions $n_{\mathrm{b}}$. For smooth snapshots, a small number of dominant modes typically captures most of the variance in the SVD, and $E_{\mathrm{proj}}$ therefore decreases exponentially as $n_{\mathrm{b}}$ increases. More specifically, Fig.~\ref{fig:uxprojerror} shows the dependence of $E_{\mathrm{proj}}$ on $n_{\mathrm{b}}$ for the steady 2D lid-driven cavity flow on a $256^2$ grid described in Sec.~\ref{setcav}. As shown in Fig.~\ref{fig:uxprojerror}, $E_{\mathrm{proj}}$ decreases approximately exponentially as $n_{\mathrm{b}}$ increases over the range approaching the error levels considered in Cases 1 and 2. At larger $n_{\mathrm{b}}$, the decrease becomes weaker and the projection error approaches a target-dependent floor. We therefore use the exponential model as an empirical approximation over the practically relevant error range:
\begin{align}
\label{exp}
E_{\mathrm{proj}}
&\le
A\exp(-\mu_{S} n_{\mathrm{b}}),
\end{align}
where $A$ and $\mu_S$ are constants.

Combining the bounds for $E_{\mathrm{enc}}$, $E_{\mathrm{sam}}$, and $E_{\mathrm{proj}}$, we obtain
\begin{equation}
\begin{aligned}
\varepsilon
\;\le\;&
\sqrt{\sum_{i=1}^{n_{\mathrm{b}}}
\left| C_i \exp(-\mu_i \chi_i) \right|^2}
+\beta\sqrt{\frac{n_{\mathrm{b}}}{N_{\mathrm{shot}}^{\mathrm{basis}}}}
\\
&\hspace{1.1em}
+A\exp(-\mu_{S} n_{\mathrm{b}}).
\end{aligned}
\end{equation}
We evaluate the three terms on the right-hand side separately and require each contribution to be bounded by $\mathcal{O}(\varepsilon)$.
This gives the following scaling relations:
\begin{eqnarray}
\label{xihiyouka}
n_{\mathrm{b}} = \mathcal{O}(\log_2(1/\varepsilon)), \quad
\chi_i = \mathcal{O}(\log_2(1/\varepsilon)),\\ \quad \nonumber
N_{\mathrm{shot}}^{\mathrm{basis}} = \mathcal{O}\!\left((1/\varepsilon)^2\log_2(1/\varepsilon) \right).
\end{eqnarray}
Since $\chi=\max_{1\le i\le n_{\mathrm{b}}}\chi_i$, Eq.~(\ref{xihiyouka}) also gives $\chi=\mathcal{O}(\log_2(1/\varepsilon))$.

\clearpage

\bibliography{sample}

\end{document}